\documentclass[11pt,a4paper]{article}

\usepackage[left=2.35cm,top=3cm,bottom=3cm,right=2.3cm]{geometry}

\usepackage{graphicx}
\usepackage[svgnames,x11names]{xcolor}
\definecolor{lightgray}{HTML}{D5D5D5}
\definecolor{lightergray}{HTML}{EEEEEE}
\usepackage{upgreek}
\usepackage{cite}
\usepackage{xcolor,colortbl}
\usepackage{slashed}

\definecolor{col}{rgb}{.4,.4,1}
\definecolor{col1}{rgb}{.4,.4,1}
\definecolor{col2}{rgb}{.4,.5,1}
\definecolor{col3}{rgb}{.4,.6,1}
\definecolor{col4}{rgb}{.4,.7,1}
\definecolor{hard}{rgb}{0.56,0.69,0.19}
\definecolor{stuff}{rgb}{0.88,0.88,0.88}
\definecolor{soft}{rgb}{0.88,0.61,0.14}
\definecolor{math1}{rgb}{0.37,0.51,0.71}
\definecolor{math2}{rgb}{0.88,0.61,0.14}
\definecolor{math3}{rgb}{0.56,0.69,0.19}
\definecolor{math4}{rgb}{0.92,0.39,0.21}
\definecolor{math5}{rgb}{0.53,0.47,0.70}
\definecolor{charge}{rgb}{0.8 0.15 0.15}

\usepackage{amsmath}
\usepackage{mathtools}
\usepackage{amsfonts}
\usepackage{amssymb}
\usepackage{slashed}
\usepackage{enumerate}
\usepackage[caption=false]{subfig}
\usepackage[colorlinks=true
,urlcolor=blue
,anchorcolor=blue
,citecolor=blue
,filecolor=blue
,linkcolor=blue
,menucolor=blue
,linktocpage=true
,pdfproducer=medialab
]{hyperref}
\numberwithin{equation}{section}

\usepackage{listings}
\usepackage{pgf}

\usepackage{xspace}

\usepackage{tikz}
\usetikzlibrary{shapes.geometric}
\usetikzlibrary{decorations.pathmorphing,positioning,decorations.pathreplacing,shapes,calc}
\usetikzlibrary{decorations.pathreplacing}
\usetikzlibrary{decorations.markings}
\usetikzlibrary{arrows}
\tikzset{
    photon/.style={decorate, decoration={snake,amplitude=1.5pt,segment length=4pt}},
    tightphoton/.style={decorate, decoration={snake,amplitude=1.5pt,segment length=5pt}},
    zigzag it/.style={decorate, decoration=zigzag},
    gluon/.style={decorate, draw=black,decoration={coil,amplitude=4pt, segment length=5pt}},
    tightgluon/.style={decorate, draw=black,decoration={coil,amplitude=2pt, segment length=3pt}}
}
\def\centerarc[#1](#2,#3)(#4:#5:#6)
    { \draw[#1] ({#2+#6*cos(#4)},{#3+#6*sin(#4)}) arc (#4:#5:#6); }

\newcommand\eik{\mathcal{E}}

\usepackage{amsmath,graphicx}

\newcommand{\wideeq}[2][1.5]{
  \mathrel{\overset{#2}{\scalebox{#1}[1]{$=$}}}
}

\catcode`,\active

\catcode`\,12

\begin{document}
\thispagestyle{empty}

\begin{flushright}
FR-PHENO-2023-04
\end{flushright}
\vspace{3em}
\begin{center}
{\Large\bf The LBK theorem to all orders}
\\
\vspace{3em}
{\sc
Tim\,Engel
}\\[2em]
{\sl Albert-Ludwigs-Universität Freiburg, Physikalisches Institut, \\
Hermann-Herder-Straße 3, D-79104 Freiburg, Germany}
\setcounter{footnote}{0}
\end{center}
\vspace{6ex}

\begin{center}
\begin{minipage}{15.3truecm}
{
We study the soft limit of one-photon radiation at next-to-leading power (NLP) in the framework of heavy-quark effective theory (HQET) to all orders in perturbation theory. We establish the soft theorem that for unpolarised scattering the radiative contribution up to NLP is entirely determined by the non-radiative amplitude. This generalises the Low-Burnett-Kroll (LBK) theorem for QED to all orders. All hard matching corrections can be calculated by applying the LBK differential operator to the non-radiative amplitude. The virtual corrections in the effective theory vanish beyond one loop, resulting in a one-loop exact soft function. As a first, non-trivial application we calculate the real-virtual-virtual electron-line corrections to muon-electron scattering at NLP in the soft limit.
}
\end{minipage}
\end{center}

\newpage

\section {Introduction}
\label{sec:intro}

Scattering amplitudes are among the main building blocks of theoretical predictions in particle phenomenology. Hence, progress in understanding their analytic structure is crucial for the Standard Model high-precision program. Particularly important in this regard is the study of the kinematic limits where external states become soft and/or collinear. In this infrared (IR) regime, the complicated multi-scale amplitudes factorise and universal, simplified structures emerge.

At leading power (LP), the soft and collinear limits capture the IR divergences of the amplitude. The corresponding factorisation theorems form the basis for the construction of IR subtraction schemes as well as for the all-order resummation of large logarithms. A significant effort has therefore been put into the study of these limits at LP. To match the experimental precision, the resummation of large logarithms beyond LP has become increasingly important in recent years. An improved understanding of the IR limits at next-to-leading power (NLP) is thus highly desirable.

In this work we focus on one-photon radiation and investigate the corresponding NLP soft limit to all orders in perturbation theory. This is to be seen in the context of high-precision theory predictions for low-energy experiments with leptons. In order to ensure reliable background predictions, the calculation of QED corrections beyond next-to-leading order (NLO) has become necessary. In addition to this phenomenological relevance, the simple Abelian nature of QED offers a clean environment to investigate fundamental properties of quantum field theory.

More than five decades ago, the seminal work of Yennie, Frautschi, and Suura (YFS)~\cite{Yennie:1961ad} demonstrated the astonishing simplicity of the LP soft limit in QED. Due to its Abelian nature, all soft virtual corrections cancel and the radiative contribution factorises into the non-radiative amplitude multiplied by a tree-level eikonal factor. As a result, all IR singularities in QED exponentiate.  In QCD, on the other hand, this is not the case since genuine loop corrections to the soft current exist~\cite{Catani:2000pi}.

For a long time, the soft NLP factorisation of one-photon radiation was only understood at tree level. In the case of unpolarised scattering, Low, Burnett, and Kroll (LBK) showed that not only the LP but also the NLP contribution is entirely determined by the non-radiative amplitude via the application of a differential operator~\cite{Low:1958sn,Burnett:1967km}. Furthermore, a similar tree-level relation is also known to exist for polarised fermions~\cite{Tarasov:1968zrq,Fearing:1973eh}. This so-called LBK theorem was later extended to the massless case in~\cite{DelDuca:1990gz} where a radiative jet function is introduced to take into account collinear effects. In recent years, extensions of the massless version of the theorem to QCD were developed both in the framework of diagrammatic factorisation~\cite{Luo:2014wea,Bonocore:2015esa,Bonocore:2016awd,Laenen:2020nrt} as well as in SCET~\cite{Larkoski:2014bxa,Beneke:2019oqx,Liu:2021mac}. Furthermore, analogous soft theorems have been studied in gravity where even at next-to-next-to-leading power a relationship to the non-radiative process exists~\cite{Cachazo:2014fwa,Bern:2014vva,Beneke:2021umj} .

Very recently, the original version of the theorem for massive QED was generalised to the one-loop level in~\cite{Engel:2021ccn}. This was done using a diagrammatic approach combined with the method of regions~\cite{Beneke:1997zp}. It was shown that at the level of the unpolarised squared amplitude the relationship to the non-radiative process is preserved. Contrary to the LP limit, however, the soft virtual corrections do not vanish. The tree-level LBK theorem is thus supplemented with a soft function that takes these effects into account. In~\cite{Kollatzsch:2022bqa} this formalism was extended to also describe polarised scattering. In this case one obtains an additional contribution at one loop that is not reducible to the non-radiative process.

These recent extensions of the LBK theorem emerged in the context of fully-differential next-to-next-to-leading order (NNLO) QED calculations. In these computations large scale hierarchies typically arise due to small but non-negligible fermion masses. As a consequence, radiative amplitudes are often hampered by numerical instabilities in the phase-space region where the emitted photon becomes simultaneously soft and collinear. The soft expansion of the corresponding amplitude at NLP can be used to achieve an accurate and stable implementation of this problematic contribution. So far, this next-to-soft stabilisation procedure has been successfully applied to Bhabha~\cite{Banerjee:2021mty}, M{\o}ller~\cite{Banerjee:2021qvi}, and muon-electron~\cite{Broggio:2022htr} scattering as well as lepton-pair production~\cite{Kollatzsch:2022bqa}.

The purpose of this work is to present the all-order generalisation of the LBK theorem for QED. We begin with a brief discussion of the tree-level version of the theorem and the corresponding proof in Section~\ref{sec:lbk_tree}. Contrary to~\cite{Engel:2021ccn}, an effective field theory (EFT) approach is employed here to describe soft photon emission. The suitable EFT in this case is the Abelian version of heavy-quark effective theory (HQET)~\cite{Eichten:1989zv,Georgi:1990um,Neubert:1993mb,Manohar:2000dt,Grozin:2004yc} also known as the Bloch-Nordsieck model~\cite{Bloch:1937pw}. In Section~\ref{sec:eft} we introduce the corresponding framework and establish the notation. Section~\ref{sec:soft} considers the general structure of the virtual corrections in the EFT and demonstrates that they vanish beyond one loop. This shows that the soft function calculated in~\cite{Engel:2021ccn} is one-loop exact. The proof is based on an extension of the conventional eikonal identity to off-shell emission, which is derived in Appendix~\ref{sec:eik_id}. Section~\ref{sec:hard} discusses the hard matching corrections and shows that they can be related to the non-radiative process in close analogy to the original tree-level proof of the LBK theorem. The complete, all-order formulation of the LBK theorem is presented in Section~\ref{sec:lbk}. A first, non-trivial application of the theorem is discussed in Section~\ref{sec:muone} in the context of muon-electron scattering. Finally, we conclude in Section~\ref{sec:conclusion}.

\section{The LBK theorem at tree level}
\label{sec:lbk_tree}

This section provides a brief introduction to the original tree-level version of the LBK theorem of~\cite{Low:1958sn,Burnett:1967km}. We restrict ourselves to a sketch of the proof here and point the reader to~\cite{Engel:2021ccn} for a more comprehensive review. We consider the soft limit of a generic QED process
\begin{equation}
	\label{eq:scattering_qed}
	\sum_{j=1}^n f_j(p_j) \to \gamma(k) \, ,
\end{equation}
with $n$ fermions $f_j$ conventionally defined as incoming and a single emitted soft photon $\gamma$. The corresponding momenta are denoted by $p_j$ and $k$, respectively. In principle, hard photons can also be present as external states. However, since they do not participate in soft interactions directly, they can be ignored in the following discussion. The photon momentum $k$ is taken be much smaller than all other scales in the process including all fermion masses $m_j$. The scale hierarchy thus reads
\begin{equation}
	\label{eq:hierarchy}
	\lambda \sim \frac{k}{p_j} \sim \frac{k}{m_j} \ll 1 \,  ,
\end{equation}
where the book-keeping parameter $\lambda$ is introduced to facilitate the power counting. We emphasize that fermions with a small or vanishing mass are not considered here. The amplitude corresponding to the process~\eqref{eq:scattering_qed} and expanded in the soft momentum up to NLP reads
\begin{equation}
	\mathcal{A}_{n+1} 
	= \frac{1}{\lambda} \mathcal{A}_{n+1}^\text{LP} 
	+  \mathcal{A}_{n+1}^\text{NLP} + \mathcal{O}(\lambda) \, .
\end{equation}
In this section we calculate this expansion generically in the case of the tree-level amplitude $\mathcal{A}_{n+1}^{(0)}$.

We split the amplitude into contributions due to external and internal emission
\begin{equation}
	\label{eq:lbk_split}
	\mathcal{A}_{n+1}^{(0)}
	=  \Bigg(\,  \sum_i \begin{tikzpicture}[scale=.8,baseline={(0,-.1)}]
        	      
   \draw[line width=.3mm]  (-1.2,0) -- (1,0);
   \draw[line width=.3mm]  (1,0) -- (1.75,.75);
   \draw[line width=.3mm]  (1,0) -- (1.75,-.75);

   \draw[line width=.3mm]  [fill=black] (1.8,.25) circle (0.02);
   \draw[line width=.3mm]  [fill=black] (1.85,0.) circle (0.02);
   \draw[line width=.3mm]  [fill=black] (1.8,-.25) circle (0.02);

   \node at (-1.1,.3) {\footnotesize{$p_i$}};

   \draw[line width=0.3mm,photon]  (-.2,0) -- (-.2,1) node[right]{\footnotesize{$k$}};

   \draw[line width=.3mm]  [fill=stuff] (1,0) circle (0.55) node[]{\footnotesize{$\Gamma_\text{ext}^{(0)}$}};
       	   \end{tikzpicture}
	  \, \Bigg)
	   +
	   \begin{tikzpicture}[scale=.8,baseline={(0,-.1)}]
        	      
   \draw[line width=.3mm]  (-.25,0) -- (1,0);
   \draw[line width=.3mm]  (1,0) -- (1.75,.75);
   \draw[line width=.3mm]  (1,0) -- (1.75,-.75);

   \draw[line width=.3mm]  [fill=black] (1.8,.25) circle (0.02);
   \draw[line width=.3mm]  [fill=black] (1.85,0.) circle (0.02);
   \draw[line width=.3mm]  [fill=black] (1.8,-.25) circle (0.02);

   \draw[line width=0.3mm,photon]  (1,0) -- (1,1) node[right]{\footnotesize{$k$}};

   \draw[line width=.3mm]  [fill=stuff] (1,0) circle (0.55) node[]{\footnotesize{$\Gamma_\text{int}^{(0)}$}};
       	   \end{tikzpicture} \,
	   =
	   \mathcal{A}_{n+1}^{(0),\text{ext}} +  \mathcal{A}_{n+1}^{(0),\text{int}}\, .
\end{equation}
All external spinors other than the one of the emitting leg are included in the definition of $\Gamma_\text{ext}^{(0)}$. Hence, the non-radiative amplitude can be written as $\mathcal{A}_n^{(0)}=\Gamma_\text{ext}^{(0)}u(p_i)$. The expansion of the external emission amplitude is given by
\begin{equation}
	\label{eq:external_exp}
	\mathcal{A}_{n+1}^{(0),\text{ext}}
	= \sum_{i=1}^n Q_i  \Big(
	-\frac{1}{\lambda} \frac{\epsilon \cdot p_i}{k \cdot p_i} \Gamma_\text{ext}^{(0)}(\{p\})
	+ \frac{\epsilon \cdot p_i}{k \cdot p_i} k \cdot \frac{\partial}{\partial p_i} \Gamma_\text{ext}^{(0)}(\{p\})
	+ \frac{\Gamma_\text{ext}^{(0)}(\{p\}) \slashed{k} \slashed{\epsilon}}{2 k \cdot p_i}
	\Big) u(p_i) + \mathcal{O}(\lambda) \, .
\end{equation}
We denote the fermion charge by $Q_i$, with $Q_i=-e<0$ for an incoming particle or an outgoing antiparticle and $Q_i=+e$ otherwise.\footnote{In~\cite{Engel:2021ccn} a different convention for the fermion charge was used where $e<0$.} Note that the set of momenta $\{p\}=\{p_1,…,p_n \}$ satisfies the radiative momentum conservation $\sum_{i=1}^n p_i = k$.

The key idea of the LBK theorem is that the non-factorisable internal emission contribution can be entirely predicted from gauge invariance. Based on the expansion~\eqref{eq:external_exp}, the split~\eqref{eq:lbk_split} can be made gauge invariant (to NLP) via the modification
\begin{equation}
	\label{eq:split_gaugeinv}
	\mathcal{A}_{n+1}^{(0)} 
	= \mathcal{A}_{n+1}^{(0),\text{ext}} +  \mathcal{A}_{n+1}^{(0),\text{int}}
	=  \mathcal{A}_{n+1}^\mathrm{I}+ \mathcal{A}_{n+1}^\mathrm{II}
\end{equation}
with
\begin{subequations}
\label{eq:lbk_internal_gaugeinv}
\begin{align}
	\mathcal{A}_{n+1}^\mathrm{I} 
	&= \epsilon_\mu \mathcal{A}_{n+1}^{\mathrm{I},\mu}
	\equiv \mathcal{A}_{n+1}^{(0),\text{ext}} - \sum_{i=1}^{n} Q_i \epsilon \cdot \frac{\partial}{\partial p_i}
	\Gamma_\text{ext}^{(0)}(\{p\}) u(p_i) \, ,
	\\
	\mathcal{A}_{n+1}^\mathrm{II} 
	&= \epsilon_\mu \mathcal{A}_{n+1}^{\mathrm{II},\mu}
	\equiv \mathcal{A}_{n+1}^{(0),\text{int}} + \sum_{i=1}^{n} Q_i \epsilon \cdot \frac{\partial}{\partial p_i}
	\Gamma_\text{ext}^{(0)}(\{p\}) u(p_i) \, .
\end{align}
\end{subequations}
Charge conservation $\sum_{i=1}^n Q_i=0$ indeed implies that $k \cdot \mathcal{A}_{n+1}^\mathrm{I} \sim \mathcal{O}(\lambda^2)$. Because the total amplitude is gauge invariant we also have $k \cdot \mathcal{A}_{n+1}^\mathrm{II} \sim \mathcal{O}(\lambda^2)$. Following~\cite{Adler:1966gc} we now show that $\mathcal{A}_{n+1}^{\mathrm{II}}$ only contributes at $ \mathcal{O}(\lambda)$ and can thus be neglected to NLP. The starting point is the observation that for an arbitrary $k$-independent four-vector function $\mathcal{A}^\mu$ the relation $k \cdot \mathcal{A} = \mathcal{O}(\lambda^2)$ directly implies $\mathcal{A}^\mu = 0$. The leading term of the internal emission amplitude $\mathcal{A}_{n+1}^{\mathrm{II},\mu}$ starts at $\mathcal{O}(\lambda^0)$. Of course, this is not equivalent to being $k$ independent. For example, terms of the form  $k^\mu/k \cdot p_j \sim \lambda^0$ could be present. The crucial observation is that at tree level internal emission does not give rise to $1/k$ poles. As a consequence, the leading $\mathcal{O}(\lambda^0)$ term of $\mathcal{A}_{n+1}^{\mathrm{II},\mu}$ is indeed independent of $k$ and has to vanish in order for $k \cdot \mathcal{A}_{n+1}^\mathrm{II} \sim \mathcal{O}(\lambda^2)$ to be satisfied.

The $\mathcal{A}_{n+1}^{\mathrm{II}}$ contribution can thus be neglected in~\eqref{eq:split_gaugeinv} and the total amplitude simplifies to
\begin{equation}
	\mathcal{A}_{n+1}^{(0)}
	= \sum_{i=1}^n Q_i  \Big(
	-\frac{1}{\lambda} \frac{\epsilon \cdot p_i}{k \cdot p_i} \Gamma_\text{ext}^{(0)}(\{p\})
	+ \frac{\Gamma_\text{ext}^{(0)}(\{p\}) \slashed{k} \slashed{\epsilon}}{2 k \cdot p_i} 
	+ [\epsilon \cdot D_i \Gamma_\text{ext}^{(0)}(\{p\})]
	\Big) u(p_i) + \mathcal{O}(\lambda)
\end{equation}
with the LBK differential operator
\begin{equation}
	D_i^\mu
	 = \frac{p_i^\mu}{k \cdot p_i} k  \cdot \frac{\partial}{\partial p_i}
	 - \frac{\partial}{\partial p_{i,\mu}} \, .
\end{equation}
At the level of the unpolarised squared amplitude we arrive at the original version of the LBK theorem
\begin{equation}
	\label{eq:lbk_original}
	\mathcal{M}_{n+1}^{(0)}(\{p\},k)
	= \sum_{i,j=1}^n Q_i Q_j \Big(
	-\frac{1}{\lambda^2} \frac{p_i \cdot p_j}{(k \cdot p_i) (k \cdot p_j)}
	+ \frac{1}{\lambda} \frac{p_j \cdot D_i}{k \cdot p_j}
	\Big) \mathcal{M}_n^{(0)}(\{p\})
	+ \mathcal{O}(\lambda^0) \, ,
\end{equation}
which shows that the radiative contribution to NLP is completely determined by the non-radiative process. This  statement also holds for polarised scattering at tree level as first shown in~\cite{Tarasov:1968zrq,Fearing:1973eh} and recently reviewed in~\cite{Kollatzsch:2022bqa}.

In~\eqref{eq:lbk_original} the non-radiative squared amplitude $\mathcal{M}_n^{(0)}$ is evaluated with a set of momenta 
that does not satisfy non-radiative momentum conservation since $\sum_i p_i = \mathcal{O}(\lambda)$. While unproblematic at tree level, this leads to complications once loop corrections are taken into account. It is therefore more convenient to express the LBK theorem in terms of kinematic invariants of the non-radiative process.\footnote{An alternative approach for massless fermions and well suited for numerical calculations can be found in~\cite{Bonocore:2021cbv}.} Following~\cite{Engel:2021ccn}, we can rewrite~\eqref{eq:lbk_original} as
\begin{equation}
	\label{eq:lbk_tree}
	\mathcal{M}_{n+1}^{(0)}(\{p\},k)
	= \sum_{i,j=1}^n Q_i Q_j\Big(
	-\frac{1}{\lambda^2} \frac{p_i \cdot p_j}{(k \cdot p_i) (k \cdot p_j)}
	+ \frac{1}{\lambda} \frac{p_j \cdot \tilde{D}_i}{k \cdot p_j}
	\Big) \mathcal{M}_n^{(0)}(\{s\},\{m^2\})
	+ \mathcal{O}(\lambda^0)
\end{equation}
with the redefined LBK operator
\begin{equation}
	\label{eq:lbkop_new}
	 \tilde{D}_i^\mu
	 =\sum_L \Big(
	 \frac{p_i^\mu}{k \cdot p_i} k \cdot \frac{\partial s_L}{\partial p_i}
	 - \frac{\partial s_L}{\partial p_{i,\mu}} \Big) 
	 \frac{\partial}{\partial s_L} \, .
\end{equation}
As emphasized in~\cite{Engel:2021ccn}, different parametrisations of the invariants $\{s\}=\{s(\{p\},\{m^2\})\}$ differ at the level of $\mathcal{O}(\lambda)$ as a consequence of the aforementioned violation of momentum conservation. For this reason, it is crucial to use the same definition both in the evaluation of the non-radiative contribution as well as in the calculation of the derivatives $\partial s_L/\partial p_i^\mu$.

Beyond tree level the above proof is not applicable anymore for two reasons. First, the photon vertex of the external emission amplitude now obtains higher-order corrections. These contributions classify as internal emission but at the same time give rise to $1/k$ poles. Second, one obtains contributions from the soft region of loop integrals where the loop momentum scales as $\ell \sim \lambda$. This interferes with the naive power counting~\eqref{eq:hierarchy}. A generalisation of the LBK theorem beyond tree level is, however, still possible. In~\cite{Engel:2021ccn} the method of regions was used to evaluate the soft contribution for generic QED processes at one loop. By doing so, it was possible to show that a factorisation into a universal soft function and the non-radiative amplitude emerges. Furthermore, also the hard region is completely determined by the non-radiative process via a direct generalisation of~\eqref{eq:lbk_tree}. This latter statement is only valid at the level of the unpolarised squared amplitude. If polarised fermions are considered one obtains additional contributions that cannot be related to the non-radiative process~\cite{Kollatzsch:2022bqa}. In the following sections, we extend this one-loop result to all orders in perturbation theory. In order to do so, we work in the framework of HQET instead of full QED, which makes the universal structure of the NLP limit more transparent.

\section{One-photon radiation at NLP in HQET}
\label{sec:eft}

The suitable EFT to describe soft photon radiation is the Abelian version of HQET. In the following we call this simpler version heavy-lepton effective theory (HLET). In this framework, fermions are taken to be nearly on shell and the corresponding momenta are decomposed as
\begin{equation}
	p_j^\mu = m_j v_j^\mu + q_j^\mu
\end{equation}
with $v_j^2=1$ and $|q_j^2| \ll m_j^2$. The large momentum component of the fermion field $\psi_j$ is given by
\begin{equation}
	h_{v_j} (x)= e^{i m_j v_j \cdot x} P_+ \psi_j(x)
\end{equation}
with the projection operator $P_+=(1+\slashed{v}_j)/2$. The heavy fermion field $h_{v_j}$ only contains field excitations corresponding to the small residual momentum $q_j$. Upon integrating out the heavy degrees of freedom, the HLET Lagrangian to NLP is given by~\cite{Eichten:1990vp,Falk:1990pz}
\begin{align}
	\label{eq:lag_hqet}
	\mathcal{L}_\text{HLET}
	 = \mathcal{L}_\text{LP} 
	 + \lambda \big( \mathcal{L}_\text{kin} + \mathcal{L}_\text{mag} \big)
	 + \mathcal{O}(\lambda^2)
\end{align}
with
\begin{subequations}
\begin{align}
	\label{eq:lag_lp}
	\mathcal{L}_\text{LP} 
	& = \sum_{j=1}^n \Big( \bar{h}_{v_j} i v_j \cdot D\, h_{v_j} \Big)-  \frac{1}{4}F_{\mu\nu} F^{\mu\nu} \, ,
	\\
	 \mathcal{L}_\text{kin} 
	 & = \sum_{j=1}^n \frac{C_\text{kin}}{2 m_j} \bar{h}_{v_j} (i D_{\perp_j})^2 h_{v_j} \, ,
	 \\
	 \label{eq:lag_mag}
	  \mathcal{L}_\text{mag}
	  & = \sum_{j=1}^n \frac{e C_\text{mag}}{4m_j} \bar{h}_{v_j} \sigma_{\mu\nu} F^{\mu\nu} h_{v_j} \, .
\end{align}
\end{subequations}
The expansion is written in terms of the book-keeping parameter $\lambda \sim |q_j^2|/m_j^2 \ll 1$.  The usual definitions are used for the covariant derivative $D^\mu = \partial^\mu - i e A^\mu$, the electromagnetic field strength $F^{\mu\nu} = \partial^\mu A^\nu - \partial^\nu A^\mu$, and $\sigma_{\mu\nu}=i[\gamma^\mu,\gamma^\nu]/2$. Furthermore, we use the notation
\begin{equation}
	g_{\perp_j}^{\mu\nu} = g^{\mu\nu}-v_j^\mu v_j^\nu
\end{equation}
for the perpendicular component of the metric w.r.t.\ the direction $v_j$ and
\begin{equation}
	p_{\perp_j}^\mu
	= g_{\perp_j}^{\mu \nu} p_\nu
	= p^\mu - v_j^\mu v_j \cdot p
\end{equation}
for any four-momentum $p$.

The HLET Lagrangian~\eqref{eq:lag_hqet} exhibits a $\mathrm{SU}(n)$ flavour symmetry for the $n$ different fermion directions $v_j$. At LP, this combines with an additional $\mathrm{SU}(2)$ spin-rotation invariance to the well-known $\mathrm{SU}(2n)$ heavy-quark spin-flavour symmetry~\cite{Isgur:1989vq,Georgi:1990um}. At NLP the spin symmetry is broken by the occurrence of the magnetic coupling of the fermion spin to the photon field in $\mathcal{L}_\text{mag}$. In addition to the remaining unbroken flavour symmetry, the NLP Lagrangian is invariant under small $\mathcal{O}(\lambda)$ changes of the directions $v_j$. This so-called reparametrisation invariance implies for the kinetic Wilson coefficient~\cite{Luke:1992cs,Chen:1993np}
\begin{equation}
	C_\text{kin} = 1
\end{equation}
to all orders. We are therefore only left with the magnetic coefficient $C_\text{mag}$ that obtains non-trivial matching corrections. These are known up to three loops~\cite{Czarnecki:1997dz,Grozin:2007fh}.

We now give the Feynman rules that are generated by the Lagrangian~\eqref{eq:lag_hqet}. All momenta are taken to be incoming and we denote the fermion charge by $Q_j$ as in the previous section. At LP we find the familiar eikonal Feynman rules
\begin{equation}
\label{eq:eikonal_rules}
	\begin{tikzpicture}[scale=.8,baseline={(0,-.1)}]

   \draw[line width=.3mm,style=double]  (-1,0) -- (1,0);
   \node at (-.8,.3) {\footnotesize{$v_j,q$}};

       \end{tikzpicture}
       = \frac{i}{q \cdot v_j + i 0} P_+
       \, , \quad\quad\quad
       	\begin{tikzpicture}[scale=.8,baseline={(0,-.1)}]

   \draw[line width=.3mm,style=double]  (-1,0) -- (1,0);
   \draw[line width=.3mm, photon] (0,0) -- (0,1) node[right] {\footnotesize{$k,\mu$}};

   \node at (-.8,.3) {\footnotesize{$v_j,q$}};

       \end{tikzpicture}
       = -i Q_j v_j^\mu\, ,
\end{equation}
where the heavy-fermion propagator follows the usual causal $+i0$ prescription. At NLP, the kinetic term $\mathcal{L}_\text{kin}$ gives rise to the three vertices
\begin{equation}
\label{eq:vertex_kin}
	\begin{tikzpicture}[scale=.8,baseline={(0,-.1)}]

   \draw[line width=.3mm,style=double]  (-1,0) -- (1,0);
   \node at (-.8,.3) {\footnotesize{$v_j,q$}};
   \node[rectangle,fill=black] at (0,0) {};

      	\end{tikzpicture}
       	=  \frac{i}{2m_j} q_{\perp_j}^2
	\, , \quad\quad
	\begin{tikzpicture}[scale=.8,baseline={(0,-.1)}]

   \draw[line width=.3mm,style=double]  (-1,0) -- (1,0);
   \node at (-.8,.3) {\footnotesize{$v_j,q$}};
   \node[rectangle,fill=black] at (0,0) {};
   \draw[line width=.3mm, photon] (0,0) -- (0,1) node[right] {\footnotesize{$k,\mu$}};

      	\end{tikzpicture}
	= \frac{-i Q_j}{2m_j} (2q+k)^\mu_{\perp_j}
	\, , \quad\quad
	\begin{tikzpicture}[scale=.8,baseline={(0,-.1)}]

   \draw[line width=.3mm,style=double]  (-1,0) -- (1,0);
   \node[rectangle,fill=black] at (0,0) {};
   \node at (-.8,.3) {\footnotesize{$v_j,q$}};
   \draw[line width=.3mm, photon] (0,0) -- (-.6,1) node[left]{\footnotesize{$\mu$}};
   \draw[line width=.3mm, photon] (0,0) -- (.6,1) node[right]{\footnotesize{$\nu$}};

      	\end{tikzpicture}
	= \frac{i Q_j^2}{m_j} g_{\perp_j}^{\mu\nu} \, .
\end{equation}
The only vertex with a non-trivial Dirac structure comes from the magnetic term  $\mathcal{L}_\text{mag}$ and reads
\begin{equation}
\label{eq:vertex_mag}
	\begin{tikzpicture}[scale=.8,baseline={(0,-.1)}]

   \draw[line width=.3mm,style=double]  (-1,0) -- (1,0);
   \node at (-.8,.3) {\footnotesize{$v_j,q$}};
   \node[draw,isosceles triangle,fill=black,rotate=-90,minimum size=.3cm,inner sep = 0pt]{};
   \draw[line width=.3mm, photon] (0,0) -- (0,1) node[right] {\footnotesize{$k,\mu$}};

      	\end{tikzpicture}
       	=  \frac{Q_j C_\text{mag}}{2m_j} \sigma_{\mu\nu} k^\nu \, .
\end{equation}

In order to describe the HLET version of the scattering process~\eqref{eq:scattering_qed}, 
\begin{equation}
	\sum_{j=1}^n h_{v_j}  \to \gamma(k)\, ,
\end{equation}
the Lagrangian~\eqref{eq:lag_hqet} has to be supplemented with the most general, gauge-invariant operators that generate this process, both at LP as well as at NLP. The LP operator, $\mathcal{O}_\text{LP}$, consists of the $n$ fermion fields only. The NLP operator, $\mathcal{O}_\text{NLP}$, contains in addition one covariant derivative which raises the mass dimension of the operator by one unit. Since we are going to evaluate the corresponding contributions in full QED in Section~\ref{sec:hard}, the exact structure of these operators does not matter in our case. Important for the following discussion is only the fact that at LP no soft photons are present in the operator, while at NLP there can be up to one photon encoded in the covariant derivative. Therefore, these operators generate the following vertices:
\begin{equation}
	\label{eq:vertices_ops}
	\begin{tikzpicture}[scale=1.,baseline={(0,-.1)}]
        		   
   \draw[line width=.3mm,style=double]  (-.25,0) -- (1,0);
   \draw[line width=.3mm,style=double]  (1,0) -- (1.75,.75);
   \draw[line width=.3mm,style=double]  (1,0) -- (1.75,-.75);

   \draw[line width=.3mm]  [fill=stuff] (1,0) circle (0.55) node[]{\footnotesize{$\mathcal{O}_\text{LP}$}};

   \draw[line width=.3mm]  [fill=black] (1.8,.25) circle (0.02);
   \draw[line width=.3mm]  [fill=black] (1.85,0.) circle (0.02);
   \draw[line width=.3mm]  [fill=black] (1.8,-.25) circle (0.02);
      	\end{tikzpicture}
	\qquad\qquad\qquad
	\begin{tikzpicture}[scale=1.,baseline={(0,-.1)}]
        		   
   \draw[line width=.3mm,style=double]  (-.25,0) -- (1,0);
   \draw[line width=.3mm,style=double]  (1,0) -- (1.75,.75);
   \draw[line width=.3mm,style=double]  (1,0) -- (1.75,-.75);

   \draw[line width=.3mm]  [fill=stuff] (1,0) circle (0.55) node[]{\footnotesize{$\mathcal{O}_\text{NLP}$}};

   \draw[line width=.3mm]  [fill=black] (1.8,.25) circle (0.02);
   \draw[line width=.3mm]  [fill=black] (1.85,0.) circle (0.02);
   \draw[line width=.3mm]  [fill=black] (1.8,-.25) circle (0.02);
      	\end{tikzpicture}
	\qquad\qquad\qquad
	\begin{tikzpicture}[scale=1.,baseline={(0,-.1)}]
        		   
   \draw[line width=.3mm,style=double]  (-.25,0) -- (1,0);
   \draw[line width=.3mm,style=double]  (1,0) -- (1.75,.75);
   \draw[line width=.3mm,style=double]  (1,0) -- (1.75,-.75);

   \draw[line width=0.3mm,photon]  (1,0) -- (1,1.1);

   \draw[line width=.3mm]  [fill=stuff] (1,0) circle (0.55) node[]{\footnotesize{$\mathcal{O}_\text{NLP}$}};

   \draw[line width=.3mm]  [fill=black] (1.8,.25) circle (0.02);
   \draw[line width=.3mm]  [fill=black] (1.85,0.) circle (0.02);
   \draw[line width=.3mm]  [fill=black] (1.8,-.25) circle (0.02);
      	\end{tikzpicture}
\end{equation}

The full NLP Lagrangian is given by
\begin{equation}
	\mathcal{L} 
	= \mathcal{L}_\text{HLET}  
	+ \mathcal{O}_\text{LP} + \mathcal{O}_\text{NLP} \, ,
\end{equation}
which naturally splits the NLP amplitude into the four contributions
\begin{equation}
\label{eq:ampNLP}
	\mathcal{A}_{n+1} 
	= \frac{1}{\lambda} \mathcal{A}_{n+1}^\text{LP}
	+ \big( \mathcal{A}_{n+1}^\text{NLP}  
	+ \mathcal{A}_{n+1}^\text{kin} 
	+ \mathcal{A}_{n+1}^\text{mag}  \big)
	+ \mathcal{O}(\lambda) \, .
\end{equation}
The LP amplitude is given by
\begin{equation}
\label{eq:ampLP}
	\mathcal{A}_{n+1}^\text{LP} 
	= \langle \gamma(k) 
	| T \big\{ \mathcal{O}_\text{LP}(0) e^{i  \int  \mathrm{d}^4 x \mathcal{L}_\text{LP}^\text{int}} \big\}
	| \prod_{j=1}^n h_{v_j} \rangle \, ,
\end{equation}
where $ \mathcal{L}_\text{LP}^\text{int}$ is the interacting part of the LP Lagrangian~\eqref{eq:lag_lp}. The NLP contribution associated with $\mathcal{O}_\text{NLP}$ is
\begin{equation}
\label{eq:ampNLPO}
	\mathcal{A}_{n+1}^\text{NLP}
	= \langle \gamma(k) 
	| T \big\{ \mathcal{O}_\text{NLP}(0) e^{i  \int  \mathrm{d}^4 x \mathcal{L}_\text{LP}^\text{int}} \big\}
	|  \prod_{j=1}^n h_{v_j} \rangle\, .
\end{equation}
The two other types of NLP contributions correspond to the combined insertions of $\mathcal{L}_\text{kin}$ or $\mathcal{L}_\text{mag}$ with $\mathcal{O}_\text{LP}$ and read
\begin{subequations}
\label{eq:ampNLPRest}
\begin{align}
        \label{eq:ampNLPkin}
	\mathcal{A}_{n+1}^\text{kin} 
	& = i\int \mathrm{d}^4 y
	\langle \gamma(k) 
	| T \big\{ \mathcal{L}_\text{kin}(y) \mathcal{O}_\text{LP}(0) e^{i  \int  \mathrm{d}^4 x \mathcal{L}_\text{LP}^\text{int}} \big\}
	|  \prod_{j=1}^n h_{v_j} \rangle \, ,
	\\
	\label{eq:ampNLPmag}
	\mathcal{A}_{n+1}^\text{mag} 
	& = i\int \mathrm{d}^4 y
	\langle \gamma(k) 
	| T \big\{ \mathcal{L}_\text{mag}(y) \mathcal{O}_\text{LP}(0) e^{i  \int  \mathrm{d}^4 x \mathcal{L}_\text{LP}^\text{int}} \big\}
	| \prod_{j=1}^n h_{v_j}  \rangle \, .
\end{align}
\end{subequations}

The following Section~\ref{sec:soft} discusses the all-order evaluation of these amplitudes. We will see that all corrections beyond one loop vanish, which yields a soft contribution that is one-loop exact. Section~\ref{sec:hard} then considers the hard matching corrections and shows that they are completely determined by the non-radiative process in the case of unpolarised scattering.

\section{The soft contribution}
\label{sec:soft}

This section presents the evaluation of the amplitude~\eqref{eq:ampNLP} to all orders in the perturbative expansion. In order to do so we rely on the conventional on-shell eikonal identity~\eqref{eq:eikid_conv} and its off-shell generalisation~\eqref{eq:eikid_gen}. Both identities are derived in Appendix~\ref{sec:eik_id}. In Section~\ref{sec:soft_LP} we prove the famous YFS result~\cite{Yennie:1961ad} that at LP all virtual soft corrections cancel. Section~\ref{sec:soft_NLP} discusses the corrections at NLP and shows that they vanish beyond one loop. In the following, all loop propagators satisfy the usual causal $+i0$ prescription.

\subsection{Leading power}
\label{sec:soft_LP}

We begin with the corrections at one and two loop and demonstrate that the sum of all contributions reduces to scaleless integrals only and thus vanishes in dimensional regularisation. Afterwards, we generalise the argument to arbitrary loop order.

The set of one-loop corrections that connect two different external legs is given by the two diagrams
\begin{subequations}
\begin{align}
    \label{eq:soft_LP_diag1_diags}
    \mathcal{A}_\text{LP}^{ij,(1)}
    &= \begin{tikzpicture}[scale=.8,baseline={(0,-.1)}]
           \draw[line width=.3mm,style=double]  (-1.2,0) -- (1,0);
   \draw[line width=.3mm,style=double]  (1,0) -- (2.7,.25);
   \draw[line width=.3mm,style=double]  (1,0) -- (2.75,0);
   \draw[line width=.3mm,style=double]  (1,0) -- (2.7,-.25);
   
   \draw[line width=.3mm]  [fill=black] (2.9,.25) circle (0.02);
   \draw[line width=.3mm]  [fill=black] (2.925,0) circle (0.02);
   \draw[line width=.3mm]  [fill=black] (2.9,-.25) circle (0.02);
   
   \node at (-1.1,.3) {\footnotesize{$v_i$}};
   \node at (2.7,-.6) {\footnotesize{$v_j$}};

   \draw[line width=0.3mm,photon]  (-.5,0) -- (-.5,1) node[right]{\footnotesize{$k$}};
   \centerarc [line width=0.3mm, photon](.5,.2)(-15:-170:1.34);

   \draw[line width=.3mm]  [fill=stuff] (1,0) circle (0.55) node[]{\footnotesize{$\mathcal{O}_\text{LP}$}};

    \node at (-.9,-.8) {\footnotesize{$\ell$}};
       \end{tikzpicture}
     + 
     \begin{tikzpicture}[scale=.8,baseline={(0,-.1)}]
          \draw[line width=.3mm,style=double]  (-1.2,0) -- (1,0);
   \draw[line width=.3mm,style=double]  (1,0) -- (2.7,.25);
   \draw[line width=.3mm,style=double]  (1,0) -- (2.75,0);
   \draw[line width=.3mm,style=double]  (1,0) -- (2.7,-.25);
   
   \draw[line width=.3mm]  [fill=black] (2.9,.25) circle (0.02);
   \draw[line width=.3mm]  [fill=black] (2.925,0) circle (0.02);
   \draw[line width=.3mm]  [fill=black] (2.9,-.25) circle (0.02);
   
   \node at (-1.1,.3) {\footnotesize{$v_i$}};
   \node at (2.7,-.6) {\footnotesize{$v_j$}};

   \draw[line width=0.3mm,photon]  (-.5,0) -- (-.5,1) node[right]{\footnotesize{$k$}};
   \centerarc [line width=0.3mm, photon](1.2,.2)(-20:-170:1.34);

   \draw[line width=.3mm]  [fill=stuff] (1,0) circle (0.55) node[]{\footnotesize{$\mathcal{O}_\text{LP}$}};

    \node at (-.2,-.8) {\footnotesize{$\ell$}};
       \end{tikzpicture}
    \\
    &= \label{eq:soft_LP_diag1}
    \int[\mathrm{d}\ell] \frac{-Q_i^2 Q_j v_i \cdot v_j \epsilon \cdot v_i}{[\ell^2][-\ell \cdot v_j]}
    \Big(
    \frac{1}{[\ell \cdot v_i][\ell \cdot v_i - k \cdot v_i]}
    +
    \frac{1}{-k \cdot v_i[\ell \cdot v_i - k \cdot v_i]}
    \Big)\mathcal{A}_{n}^\text{LP}\, ,
\end{align}
\end{subequations}
where 
\begin{equation}
     \mathcal{A}_n^{\text{LP}}
      =
      \langle 0 
	|  \mathcal{O}_\text{LP}(0)
	| \prod_{j=1}^n h_{v_j} \rangle
    = \begin{tikzpicture}[scale=.8,baseline={(0,-.1)}]
           
   \draw[line width=.3mm,style=double]  (-.25,0) -- (1,0);
   \draw[line width=.3mm,style=double]  (1,0) -- (1.75,.75);
   \draw[line width=.3mm,style=double]  (1,0) -- (1.75,-.75);

   \draw[line width=.3mm]  [fill=stuff] (1,0) circle (0.55) node[]{\footnotesize{$\mathcal{O}_\text{LP}$}};

   \draw[line width=.3mm]  [fill=black] (1.8,.25) circle (0.02);
   \draw[line width=.3mm]  [fill=black] (1.85,0.) circle (0.02);
   \draw[line width=.3mm]  [fill=black] (1.8,-.25) circle (0.02);
       \end{tikzpicture}
     = \mathcal{A}_n
\end{equation}
and the $d$-dimensional loop measure
\begin{equation}
	 [\mathrm{d}\ell] = i  \mu^{2\epsilon} \frac{\mathrm{d}^d \ell}{(2\pi)^d}
\end{equation}
absorbs the $i$ from the photon propagator. Since no external soft scale is present in $\mathcal{A}_n^\text{LP}$ it exactly corresponds to the full QED amplitude $\mathcal{A}_n$. After partial fraction decomposition
\begin{equation}\label{eq:partial_fraction_simple}
    \frac{1}{[\ell \cdot v_i][\ell \cdot v_i - k \cdot v_i]}
    = \frac{1}{-k \cdot v_i}\Big( 
    \frac{1}{[\ell \cdot v_i]}-\frac{1}{[\ell \cdot v_i - k \cdot v_i]} \Big)
\end{equation}
the amplitude~\eqref{eq:soft_LP_diag1} simplifies to
\begin{equation} \label{eq:soft_LP_diag1_simp}
    \mathcal{A}_\text{LP}^{ij,(1)}
    = \frac{Q_i \epsilon \cdot v_i}{-k \cdot v_i} \int[\mathrm{d}\ell] \frac{-Q_i Q_j v_i \cdot v_j}{[\ell^2][\ell \cdot v_i][-\ell \cdot v_j]}\, .
\end{equation}
As alluded to above, the resulting loop integral is scaleless and thus $\mathcal{A}_\text{LP}^{ij,(1)}=0$.

At the two-loop order we obtain the six contributions
\begin{align}
\label{eq:soft_LP_2L_diag1_diags}
    \mathcal{A}_\text{LP}^{ijk,(2)}
    &= \Bigg( \begin{tikzpicture}[scale=.8,baseline={(0,-.1)}]
           \draw[line width=.3mm,style=double]  (-1.2,0) -- (1,0);
   \draw[line width=.3mm,style=double]  (1,0) -- (2.7,.25);
   \draw[line width=.3mm,style=double]  (1,0) -- (2.75,0);
   \draw[line width=.3mm,style=double]  (1,0) -- (2.7,-.25);
   
   \draw[line width=.3mm]  [fill=black] (2.9,.25) circle (0.02);
   \draw[line width=.3mm]  [fill=black] (2.925,0) circle (0.02);
   \draw[line width=.3mm]  [fill=black] (2.9,-.25) circle (0.02);
   
   \node at (-1.1,.3) {\footnotesize{$v_i$}};
   \node at (2.7,-.6) {\footnotesize{$v_j$}};
   \node at (2.7,.6) {\footnotesize{$v_k$}};

   \centerarc [line width=0.3mm, photon](.75,-.9)(40:148:1.7);
    \centerarc [line width=0.3mm, photon](.5,.5)(-25:-160:1.6);
    \draw[fill=white,draw=none] (-.3,.5) circle (0.2);
     \draw[line width=0.3mm,photon]  (-.3,0) -- (-.3,1) node[left]{\footnotesize{$k$}};
  
   \draw[line width=.3mm]  [fill=stuff] (1,0) circle (0.55) node[]{\footnotesize{$\mathcal{O}_\text{LP}$}};

    \node at (0,-.7) {\footnotesize{$\ell_1$}};
    \node at (.8,1.15) {\footnotesize{$\ell_2$}};
       \end{tikzpicture}
     + 
     \begin{tikzpicture}[scale=.8,baseline={(0,-.1)}]
        
   \draw[line width=.3mm,style=double]  (-1.2,0) -- (1,0);
   \draw[line width=.3mm,style=double]  (1,0) -- (2.7,.25);
   \draw[line width=.3mm,style=double]  (1,0) -- (2.75,0);
   \draw[line width=.3mm,style=double]  (1,0) -- (2.7,-.25);
   
   \draw[line width=.3mm]  [fill=black] (2.9,.25) circle (0.02);
   \draw[line width=.3mm]  [fill=black] (2.925,0) circle (0.02);
   \draw[line width=.3mm]  [fill=black] (2.9,-.25) circle (0.02);
   
   \node at (-1.1,.3) {\footnotesize{$v_i$}};
   \node at (2.7,-.6) {\footnotesize{$v_j$}};
   \node at (2.7,.6) {\footnotesize{$v_k$}};

   \centerarc [line width=0.3mm, photon](1.25,-.7)(39:150:1.5);
    \centerarc [line width=0.3mm, photon](.5,.5)(-25:-160:1.6);
     \draw[line width=0.3mm,photon]  (-.3,0) -- (-.3,1) node[left]{\footnotesize{$k$}};
  
   \draw[line width=.3mm]  [fill=stuff] (1,0) circle (0.55) node[]{\footnotesize{$\mathcal{O}_\text{LP}$}};

    \node at (0,-.7) {\footnotesize{$\ell_1$}};
    \node at (.8,1.15) {\footnotesize{$\ell_2$}};

       \end{tikzpicture}
         + 
     \begin{tikzpicture}[scale=.8,baseline={(0,-.1)}]
        
   \draw[line width=.3mm,style=double]  (-1.2,0) -- (1,0);
   \draw[line width=.3mm,style=double]  (1,0) -- (2.7,.25);
   \draw[line width=.3mm,style=double]  (1,0) -- (2.75,0);
   \draw[line width=.3mm,style=double]  (1,0) -- (2.7,-.25);
   
   \draw[line width=.3mm]  [fill=black] (2.9,.25) circle (0.02);
   \draw[line width=.3mm]  [fill=black] (2.925,0) circle (0.02);
   \draw[line width=.3mm]  [fill=black] (2.9,-.25) circle (0.02);
   
   \node at (-1.1,.3) {\footnotesize{$v_i$}};
   \node at (2.7,-.6) {\footnotesize{$v_j$}};
   \node at (2.7,.6) {\footnotesize{$v_k$}};

   \centerarc [line width=0.3mm, photon](1.4,-.7)(39:150:1.5);
    \centerarc [line width=0.3mm, photon](1.25,.5)(-30:-160:1.5);
     \draw[line width=0.3mm,photon]  (-.3,0) -- (-.3,1) node[left]{\footnotesize{$k$}};
  
   \draw[line width=.3mm]  [fill=stuff] (1,0) circle (0.55) node[]{\footnotesize{$\mathcal{O}_\text{LP}$}};

    \node at (0,-.8) {\footnotesize{$\ell_1$}};
    \node at (.8,1.1) {\footnotesize{$\ell_2$}};

       \end{tikzpicture}\Bigg)
    + (\ell_1 \leftrightarrow \ell_2)\, .
\end{align}
Analogously to the one-loop case partial fraction decomposition reveals various cancellations among these diagrams and the amplitude reduces to the scaleless integrals
\begin{align}\label{eq:ampLP_2L}
    \mathcal{A}_\text{LP}^{ijk,(2)}
     &= \frac{Q_i \epsilon \cdot k}{-k \cdot v_i}
     \int[\mathrm{d}\ell_1] \frac{-Q_i Q_j v_i \cdot v_j}{[\ell_1][\ell_1 \cdot v_i][-\ell_1 \cdot v_j]}
     \int[\mathrm{d}\ell_2] \frac{-Q_i Q_k v_i \cdot v_k}{[\ell_2][\ell_2 \cdot v_i][-\ell_2 \cdot v_k]} \mathcal{A}_n
     = 0 \, .
\end{align}

The cancellations in~\eqref{eq:soft_LP_diag1_diags} and \eqref{eq:soft_LP_2L_diag1_diags} are special cases of the on-shell eikonal identity~\eqref{eq:eikid_conv} proven in Appendix~\ref{sec:eikid_conv}. This identity applies to an arbitrary number of photons and thus allows for a generalisation to the all-order contribution
\begin{equation}\label{eq:amp_soft_allorder}
    \mathcal{A}_\text{LP}^{i}
    = \sum_{u} \begin{tikzpicture}[scale=.8,baseline={(0,-.1)}]
           \draw[line width=.3mm,style=double]  (-1.2,0) -- (1,0);
   \draw[line width=.3mm,style=double]  (1,0) -- (2.7,.25);
   \draw[line width=.3mm,style=double]  (1,0) -- (2.75,0);
   \draw[line width=.3mm,style=double]  (1,0) -- (2.7,-.25);
   
   \draw[line width=.3mm]  [fill=black] (2.9,.25) circle (0.02);
   \draw[line width=.3mm]  [fill=black] (2.925,0) circle (0.02);
   \draw[line width=.3mm]  [fill=black] (2.9,-.25) circle (0.02);
   
   \node at (-1.1,.3) {\footnotesize{$v_i$}};

   \draw[line width=0.3mm,photon]  (-.3,0) -- (-.3,1) node[right]{\footnotesize{$k$}};
   \centerarc [line width=0.3mm, dashed](.95,.2)(0:-170:1.32);

   \draw[line width=.3mm]  [fill=white] (-.3,0) circle (0.2);
   \draw[line width=.3mm] (-.3-.15,0.-.15) -- (-.3+.15,0.+.15);
   \draw[line width=.3mm] (-.3+.15,0.-.15) -- (-.3-.15,0.+.15);

   \draw[line width=.3mm]  [fill=white] (2.25,0) circle (0.25);
   \draw[line width=.3mm] (2.25-.15,-.2) -- (2.25+.15,.2);
   \draw[line width=.3mm] (2.25+.15,-.2) -- (2.25-.15,.2);

   \draw[line width=.3mm]  [fill=stuff] (1,0) circle (0.55) node[]{\footnotesize{$\mathcal{O}_\text{LP}$}};

   \node at (.45,-1.2) {\tiny{$u$}};
   \node at (2.3,.55) {\footnotesize{$\Gamma_{n-1}$}};
       \end{tikzpicture}\,\,\,\, ,
\end{equation}
where the sum goes from zero to infinity and the dashed line represents $u$ soft photons. As in~\eqref{eq:eikid_conv} the crossed vertex represents all possible attachments of the photons. Applying the eikonal identity~\eqref{eq:eikid_conv} we find the compact expression
\begin{equation}
	\label{eq:eik_expr}
    \mathcal{A}_\text{LP}^i
    =  \frac{Q_i \epsilon \cdot v_i}{-k \cdot v_i}
    \sum_{u}
     \frac{1}{u!}
      \prod_{l=1}^u \int[\mathrm{d}\ell_l] \frac{-Q_i v_{i,\mu_l}}{[\ell_l^2] [\ell_l \cdot v_i]}
     \Gamma_{n-1}^{\mu_1…\mu_u} \mathcal{A}_{n} \, .
\end{equation}
Due to the trivial Dirac structure of the eikonal vertex~\eqref{eq:eikonal_rules} the entire soft contribution factorises. This also holds for the part of the diagram that corresponds to the non-emitting legs and justifies the factorisation into $\Gamma_{n-1}$ and the non-radiative amplitude $\mathcal{A}_n$. Furthermore, there is a double counting of contributions since both ends of the dashed lines end in a crossed vertex. The combinatorial factor $u!$ in~\eqref{eq:eik_expr} removes this double counting. All loop integrals in~\eqref{eq:eik_expr} have the same form as in~\eqref{eq:soft_LP_diag1_simp} and \eqref{eq:ampLP_2L} and are thus scaleless. This is in particular true since also $\Gamma_{n-1}$ only gives rise to scaleless linear propagators. For example, in the case where one ($u=1$) or two ($u=2$) photons connect to $\Gamma_{n-1}$ we have
\begin{equation}
	\label{eq:diag_rest}
	\Gamma_{n-1}^{\mu_1} = \sum_{j \neq i} \frac{Q_j v_j^{\mu_1}}{[-\ell_1 \cdot v_j]}
\end{equation}
and
\begin{equation}\label{eq:gamma_2photons}
	\Gamma_{n-1}^{\mu_1\mu_2} = 
     \sum_{j k \neq i}
     \frac{Q_j v_j^{\mu_1}}{[-\ell_1 \cdot v_j]} 
     \frac{Q_k v_k^{\mu_2}}{[-\ell_2 \cdot v_k]} \, .
\end{equation}
The two loop momenta factorise in~\eqref{eq:gamma_2photons} due to the eikonal identity~\eqref{eq:eikid_conv}.

In addition to the contributions~\eqref{eq:amp_soft_allorder} that connect two different external legs, single-leg corrections have to be taken into account as well. The amplitude~\eqref{eq:amp_soft_allorder} can be extended accordingly by inserting one-particle irreducible (1PI) corrections at the crossed vertices.\footnote{The general case follows from repeating the argument for multiple 1PI insertions.} In this case, the external photon and the photons represented by the dashed line attach in all possible ways to the 1PI diagram.\footnote{If the insertion was not 1PI, some attachments would result in external self-energy corrections.} Following~\eqref{eq:insertsum} this summation results in the decoupling of the 1PI contribution and renders the corresponding loop integrals scaleless. As a consequence, the generalised amplitude reduces to $\mathcal{A}_\text{LP}^i$, which in turn further simplifies to the tree-level contribution.

This proves that there are no soft corrections to the eikonal approximation in QED. Summing over all possible emitters $v_i$, the full LP amplitude is given by the tree-level expression
\begin{equation}
	\label{eq:diag_lp}
    \mathcal{A}_{n+1}^\text{LP}
    = \sum_{i=1}^n \begin{tikzpicture}[scale=.8,baseline={(0,-.1)}]
           
   \draw[line width=.3mm,style=double]  (-1.2,0) -- (1,0);
   \draw[line width=.3mm,style=double]  (1,0) -- (1.75,.75);
   \draw[line width=.3mm,style=double]  (1,0) -- (1.75,-.75);

   \draw[line width=.3mm]  [fill=black] (1.8,.25) circle (0.02);
   \draw[line width=.3mm]  [fill=black] (1.85,0.) circle (0.02);
   \draw[line width=.3mm]  [fill=black] (1.8,-.25) circle (0.02);

   \node at (-1.1,.3) {\footnotesize{$v_i$}};

   \draw[line width=0.3mm,photon]  (-.2,0) -- (-.2,+1) node[right]{\footnotesize{$k$}};

   \draw[line width=.3mm]  [fill=stuff] (1,0) circle (0.55) node[]{\footnotesize{$\mathcal{O}_\text{LP}$}};
       \end{tikzpicture} \, 
     = \sum_{i=1}^n \frac{Q_i \epsilon \cdot v_i}{-k \cdot v_i} \mathcal{A}_n \, .
\end{equation}

\subsection{Next-to-leading power}
\label{sec:soft_NLP}

The LP proof of Section~\ref{sec:soft_LP} can also be applied to the $\mathcal{O}_\text{NLP}$ contribution~\eqref{eq:ampNLPO}. Also in this case all virtual corrections vanish and the amplitude simplifies to the tree-level diagrams
\begin{align}
	\label{eq:diag_nlp}
    \mathcal{A}_{n+1}^\text{NLP}
    = \Bigg( \sum_{i=1}^n
     \begin{tikzpicture}[scale=.8,baseline={(0,-.1)}]
           
   \draw[line width=.3mm,style=double]  (-1.2,0) -- (1,0);
   \draw[line width=.3mm,style=double]  (1,0) -- (1.75,.75);
   \draw[line width=.3mm,style=double]  (1,0) -- (1.75,-.75);

   \draw[line width=.3mm]  [fill=black] (1.8,.25) circle (0.02);
   \draw[line width=.3mm]  [fill=black] (1.85,0.) circle (0.02);
   \draw[line width=.3mm]  [fill=black] (1.8,-.25) circle (0.02);

   \draw[line width=0.3mm,photon]  (-.2,0) -- (-.2,1) node[right]{\footnotesize{$k$}};

   \node at (-1.1,.3) {\footnotesize{$v_i$}};

   \draw[line width=.3mm]  [fill=stuff] (1,0) circle (0.55) node[]{\footnotesize{$\mathcal{O}_\text{NLP}$}};
       \end{tikzpicture}
      \, \,  \Bigg)
       +
       \begin{tikzpicture}[scale=.8,baseline={(0,-.1)}]
           
   \draw[line width=.3mm,style=double]  (-.25,0) -- (1,0);
   \draw[line width=.3mm,style=double]  (1,0) -- (1.75,.75);
   \draw[line width=.3mm,style=double]  (1,0) -- (1.75,-.75);

   \draw[line width=.3mm]  [fill=black] (1.8,.25) circle (0.02);
   \draw[line width=.3mm]  [fill=black] (1.85,0.) circle (0.02);
   \draw[line width=.3mm]  [fill=black] (1.8,-.25) circle (0.02);

   \draw[line width=0.3mm,photon]  (1,0) -- (1,1) node[right]{\footnotesize{$k$}};

   \draw[line width=.3mm]  [fill=stuff] (1,0) circle (0.55) node[]{\footnotesize{$\mathcal{O}_\text{NLP}$}};
       \end{tikzpicture} \,\,\,\, ,
\end{align}
where the two terms on the r.h.s.\ emerge from the two NLP vertices of~\eqref{eq:vertices_ops}. A different approach has to be taken for the two remaining NLP amplitudes, $\mathcal{A}^\text{kin}_{n+1}$ and $\mathcal{A}^\text{mag}_{n+1}$, defined in~\eqref{eq:ampNLPRest}. The insertion of the NLP vertices~\eqref{eq:vertex_kin} and~\eqref{eq:vertex_mag} gives rise to emission from off-shell heavy-fermion propagators. As a consequence, the conventional on-shell eikonal identity does not apply anymore. Instead, we rely on a generalised version of this identity derived in Appendix~\ref{sec:eikid_gen}. This yields a non-vanishing soft contribution which, however, turns out to be one-loop exact.

Both amplitudes, $\mathcal{A}_{n+1}^\text{kin}$ and $\mathcal{A}_{n+1}^\text{mag}$, receive contributions of the form
\begin{equation}
	\label{eq:diag_gen1}
	\mathcal{A}_{wrq,F}^{uts,i}
	= 
	\begin{tikzpicture}[scale=.8,baseline={(0,-.1)}]
           \draw[line width=.3mm,style=double]  (-3,0) -- (1,0);
   \draw[line width=.3mm,style=double]  (1,0) -- (2.7,.25);
   \draw[line width=.3mm,style=double]  (1,0) -- (2.75,0);
   \draw[line width=.3mm,style=double]  (1,0) -- (2.7,-.25);

   \draw[line width=.3mm]  [fill=black] (2.9,.25) circle (0.02);
   \draw[line width=.3mm]  [fill=black] (2.925,0) circle (0.02);
   \draw[line width=.3mm]  [fill=black] (2.9,-.25) circle (0.02);
   
   \node at (-2.9,.3) {\footnotesize{$v_i$}};

   \centerarc [line width=0.3mm, dashed](-1.25,0)(0:180:1);
   \centerarc [line width=0.3mm, dashed](.05,.5)(-10:-170:2.35);
   \centerarc [line width=0.3mm, dashed](-.75,0)(0:-180:.5);
   \centerarc [line width=0.3mm, dashed](.5,.3)(-10:-170:1.75);
   \centerarc [line width=0.3mm, dashed](1.,.2)(-10:-170:1.15);
   \draw [line width=0.3mm,dashed,rotate around={-135:(0,0)}] (-.4,-.2) ellipse(0.5 and 0.27);

   \draw[line width=.3mm]  [fill=white] (-2.25,0) circle (0.2);
   \draw[line width=.3mm] (-2.25-.15,0.-.15) -- (-2.25+.15,0.+.15);
   \draw[line width=.3mm] (-2.25+.15,0.-.15) -- (-2.25-.15,0.+.15);

   \draw[line width=.3mm]  [fill=black] (-1.25,0) circle (0.15);
   \node at (-1,.35) {\footnotesize{$F$}};
   \draw[line width=0.3mm,photon]  (-1.25,0) -- (-1.25,1.35) node[left]{\footnotesize{$k$}};

   \draw[line width=.3mm]  [fill=white] (-.25,0) circle (0.2);
   \draw[line width=.3mm] (-.25-.15,0.-.15) -- (-.25+.15,0.+.15);
   \draw[line width=.3mm] (-.25+.15,0.-.15) -- (-.25-.15,0.+.15);

   \draw[line width=.3mm]  [fill=white] (2.25,0) circle (0.25);
   \draw[line width=.3mm] (2.25-.15,-.2) -- (2.25+.15,.2);
   \draw[line width=.3mm] (2.25+.15,-.2) -- (2.25-.15,.2);
   \draw[line width=.3mm]  [fill=stuff] (1,0) circle (0.55) node[]{\footnotesize{$\mathcal{O}_\text{LP}$}};

   \node at (-2.1,-1) {\tiny{$w$}};
   \node at (.45,-1.2) {\tiny{$s-q$}};
   \node at (-1.7,.7) {\tiny{$r$}};
   \node at (-.5,-.65) {\tiny{$q$}};
   \node at (1.,-.7) {\tiny{$u-w$}};
   \node at (.3,1) {\tiny{$t-r$}};
   \node at (2.3,.55) {\footnotesize{$\Gamma_{n-1}$}};
       \end{tikzpicture}\,\,\,\, ,
\end{equation}
where the black dot denoted by $F$ corresponds either to one of the kinetic vertices~\eqref{eq:vertex_kin} or the magnetic vertex~\eqref{eq:vertex_mag}. The total number of photons connecting to $F$ is $s+1$ where $s$ is fixed by the specific vertex type. Any contribution from the first kinetic vertex of~\eqref{eq:vertex_kin} which only corrects the heavy-fermion propagator is thus excluded. It only enters in the second type of contribution discussed below. All dashed lines are omitted that directly result in scaleless integrals after applying the on-shell eikonal identity~\eqref{eq:eikid_conv}. Finally note that we have defined the dashed lines in pairs with a fixed combined number of photons, \emph{e.g.}\ $w+(u-w)=u$. This yields a decomposition of the amplitude into gauge-invariant subsets at each given order in perturbation theory.

For the two crossed vertices that attach to the external legs in~\eqref{eq:diag_gen1} we can use the on-shell eikonal identity~\eqref{eq:eikid_conv}. In the case of the crossed vertex that corrects the off-shell leg the generalised identity~\eqref{eq:eikid_gen} applies with
\begin{equation}
	\widetilde{\sum}_j \ell_j \cdot v_i
	\mathrel{\widehat{=}}
	\sum_{l=w+1}^{u} \ell_l \cdot v_i + \sum_{l=1}^{t+q} \tilde{\ell}_l \cdot v_i
	\, , \quad\quad\quad
	\tilde{p}  \cdot v_i
	\mathrel{\widehat{=}}
	\sum_{l=1}^{w} \ell_l \cdot v_i + \sum_{l=u+1}^{u+s-q} \ell_l \cdot v_i - k \cdot v_i \, ,
\end{equation}
where we use $\tilde{\ell}_l$ for the photon lines that only connect to the emitting leg labelled by $v_i$ and $\ell_l$ otherwise. This allows us to write the diagram~\eqref{eq:diag_gen1} as\footnote{The factorisation into the non-radiative amplitude $\mathcal{A}_n$ does not hold for the magnetic vertex~\eqref{eq:vertex_mag} due to the non-trivial Dirac structure $\sim \sigma^{\mu\nu}$. Instead, $\mathcal{A}_n$ is modified by inserting $\sigma^{\mu\nu}$ in the emitting fermion line. Nevertheless, we do not treat this case separately because the following arguments also apply here.}
\begin{equation}
\label{eq:ampgen1}
	\mathcal{A}_{wrq,F}^{uts,i} 
	=  \prod_{l=1}^{u+s-q} \int [\mathrm{d} \ell_l] \prod_{l=1}^{t+q} \int [\mathrm{d} \tilde{\ell}_l] \,
	(-1)^s S^{wu} S^{rt} 
	F_{\nu_1 … \nu_s\nu}^{wrq} 
	\epsilon^\nu \mathcal{P}_{utsq,\mu_1…\mu_{u}}^{\nu_1 … \nu_q}
	 \Gamma_{n-1}^{\nu_{q+1}…\nu_s\mu_1…\mu_{u} }\mathcal{A}_{n}
\end{equation}
with the propagator structure
\begin{align}
\label{eq:prop_structure}
\begin{split}
	 \mathcal{P}_{utsq,\mu_1…\mu_u}^{\nu_1 … \nu_q}
	 = &\Bigg( \prod_{l=1}^{u}  \frac{Q_i v_{i,\mu_l}}{[\ell_l^2] [\ell_l \cdot v_i]} \Bigg)
	\Bigg( \prod_{l=u+1}^{u+s-q}  \frac{1}{[\ell_l^2]} \Bigg)
	\Bigg( \prod_{l=1}^{q} \frac{Q_i v_i^{\nu_i}}{[\tilde{\ell}_l^2] [-\tilde{\ell}_l \cdot v_i]}  \Bigg)
	\Bigg( \prod_{l=q+1}^{t+q} \frac{Q_i^2}{[\tilde{\ell}_l^2] [\tilde{\ell}_l \cdot v_i] [-\tilde{\ell}_l \cdot v_i]} \Bigg)
	 \\ & \quad \times
	 \frac{i}{\sum_{l=1}^{u+s-q} \ell_l \cdot v_i + \sum_{l=1}^{t+q} \tilde{\ell}_l \cdot v_i - k \cdot v_i}\, .
\end{split}
\end{align}
The vertex factor $F_{\nu_1 … \nu_s\nu}^{wrq} $ corresponds to the Feynman rule associated to $F$. Furthermore, we have defined
\begin{equation}
	S^{xy} = \frac{(-1)^{x}}{x! (y-x)!}
\end{equation}
that takes into account the double counting as well as the signs from the two dashed lines with $x$ and $y-x$ photons. This includes the signs coming from~\eqref{eq:eikid_gen} as well as the $-1$ from the photon propagator.

In addition to~\eqref{eq:diag_gen1} we also have the two types of diagrams where the external photon does not connect to $F$ but instead to one of the crossed vertices of the emitting leg. In this case we can write the sum of the two contributions as
\begin{subequations}
\label{eq:ampgen2}
\begin{align}
        \label{eq:ampgen2_diags}
	\tilde{\mathcal{A}}_{wrq,F}^{uts,i}
	&= 
	\begin{tikzpicture}[scale=.8,baseline={(0,-.1)}]
           \draw[line width=.3mm,style=double]  (-3,0) -- (1,0);
   \draw[line width=.3mm,style=double]  (1,0) -- (2.7,.25);
   \draw[line width=.3mm,style=double]  (1,0) -- (2.75,0);
   \draw[line width=.3mm,style=double]  (1,0) -- (2.7,-.25);

   \draw[line width=.3mm]  [fill=black] (2.9,.25) circle (0.02);
   \draw[line width=.3mm]  [fill=black] (2.925,0) circle (0.02);
   \draw[line width=.3mm]  [fill=black] (2.9,-.25) circle (0.02);
   
   \node at (-2.9,.3) {\footnotesize{$v_i$}};

   \centerarc [line width=0.3mm, dashed](-1.25,0)(0:180:1);
   \centerarc [line width=0.3mm, dashed](.05,.5)(-10:-170:2.35);
   \centerarc [line width=0.3mm, dashed](-.75,0)(0:-180:.5);
   \centerarc [line width=0.3mm, dashed](.5,.3)(-10:-170:1.75);
   \centerarc [line width=0.3mm, dashed](1.,.2)(-10:-170:1.15);
   \draw [line width=0.3mm,dashed,rotate around={-135:(0,0)}] (-.4,-.2) ellipse(0.5 and 0.27);

   \draw[line width=0.3mm,photon]  (-2.25,0) -- (-2.25,1.35) node[left]{\footnotesize{$k$}};  
   \draw[line width=.3mm]  [fill=white] (-2.25,0) circle (0.2);
   \draw[line width=.3mm] (-2.25-.15,0.-.15) -- (-2.25+.15,0.+.15);
   \draw[line width=.3mm] (-2.25+.15,0.-.15) -- (-2.25-.15,0.+.15);

   \draw[line width=.3mm]  [fill=black] (-1.25,0) circle (0.15);
   \node at (-1,.35) {\footnotesize{$F$}};

   \draw[line width=.3mm]  [fill=white] (-.25,0) circle (0.2);
   \draw[line width=.3mm] (-.25-.15,0.-.15) -- (-.25+.15,0.+.15);
   \draw[line width=.3mm] (-.25+.15,0.-.15) -- (-.25-.15,0.+.15);

   \draw[line width=.3mm]  [fill=white] (2.25,0) circle (0.25);
   \draw[line width=.3mm] (2.25-.15,-.2) -- (2.25+.15,.2);
   \draw[line width=.3mm] (2.25+.15,-.2) -- (2.25-.15,.2);
   \draw[line width=.3mm]  [fill=stuff] (1,0) circle (0.55) node[]{\footnotesize{$\mathcal{O}_\text{LP}$}};

   \node at (-2.1,-1) {\tiny{$w$}};
   \node at (.45,-1.2) {\tiny{$s-q$}};
   \node at (-1.7,.7) {\tiny{$r$}};
   \node at (-.5,-.65) {\tiny{$q$}};
   \node at (1.,-.7) {\tiny{$u-w$}};
   \node at (.3,1) {\tiny{$t-r$}};
   \node at (2.3,.55) {\footnotesize{$\Gamma_{n-1}$}};
       \end{tikzpicture}
       +
       	\begin{tikzpicture}[scale=.8,baseline={(0,-.1)}]
           \draw[line width=.3mm,style=double]  (-3,0) -- (1,0);
   \draw[line width=.3mm,style=double]  (1,0) -- (2.7,.25);
   \draw[line width=.3mm,style=double]  (1,0) -- (2.75,0);
   \draw[line width=.3mm,style=double]  (1,0) -- (2.7,-.25);

   \draw[line width=.3mm]  [fill=black] (2.9,.25) circle (0.02);
   \draw[line width=.3mm]  [fill=black] (2.925,0) circle (0.02);
   \draw[line width=.3mm]  [fill=black] (2.9,-.25) circle (0.02);
   
   \node at (-2.9,.3) {\footnotesize{$v_i$}};

   \centerarc [line width=0.3mm, dashed](-1.25,0)(0:180:1);
   \centerarc [line width=0.3mm, dashed](.05,.5)(-10:-170:2.35);
   \centerarc [line width=0.3mm, dashed](-.75,0)(0:-180:.5);
   \centerarc [line width=0.3mm, dashed](.5,.3)(-10:-170:1.75);
   \centerarc [line width=0.3mm, dashed](1.,.2)(-10:-170:1.15);
   \draw [line width=0.3mm,dashed,rotate around={-135:(0,0)}] (-.4,-.2) ellipse(0.5 and 0.27);

   \draw[line width=.3mm]  [fill=white] (-2.25,0) circle (0.2);
   \draw[line width=.3mm] (-2.25-.15,0.-.15) -- (-2.25+.15,0.+.15);
   \draw[line width=.3mm] (-2.25+.15,0.-.15) -- (-2.25-.15,0.+.15);

   \draw[line width=.3mm]  [fill=black] (-1.25,0) circle (0.15);
   \node at (-1,.35) {\footnotesize{$F$}};

   \draw[line width=0.3mm,photon]  (-.25,0) -- (-.25,1.35) node[left]{\footnotesize{$k$}};
   \draw[line width=.3mm]  [fill=white] (-.25,0) circle (0.2);
   \draw[line width=.3mm] (-.25-.15,0.-.15) -- (-.25+.15,0.+.15);
   \draw[line width=.3mm] (-.25+.15,0.-.15) -- (-.25-.15,0.+.15);

   \draw[line width=.3mm]  [fill=white] (2.25,0) circle (0.25);
   \draw[line width=.3mm] (2.25-.15,-.2) -- (2.25+.15,.2);
   \draw[line width=.3mm] (2.25+.15,-.2) -- (2.25-.15,.2);
   \draw[line width=.3mm]  [fill=stuff] (1,0) circle (0.55) node[]{\footnotesize{$\mathcal{O}_\text{LP}$}};

   \node at (-2.1,-1) {\tiny{$w$}};
   \node at (.45,-1.2) {\tiny{$s-q$}};
   \node at (-1.7,.7) {\tiny{$r$}};
   \node at (-.5,-.65) {\tiny{$q$}};
   \node at (1.,-.7) {\tiny{$u-w$}};
   \node at (.3,1) {\tiny{$t-r$}};
   \node at (2.3,.55) {\footnotesize{$\Gamma_{n-1}$}};
       \end{tikzpicture}
       \\
        \label{eq:ampgen2_expr}
       &= 
        \prod_{l=1}^{u+s-q} \int [\mathrm{d} \ell_l] \prod_{l=1}^{t+q} \int [\mathrm{d} \tilde{\ell}_l] \,
        (-1)^s S^{wu} S^{rt} 
       \frac{Q_i \epsilon \cdot v_i}{-k \cdot v_i}
        \tilde{F}_{\nu_1 … \nu_s}^{wrq} \mathcal{P}_{utsq,\mu_1…\mu_u}^{\nu_{1} … \nu_{q}}
       \Gamma_{n-1}^{\nu_{q+1}…\nu_s\mu_1…\mu_{u}}\mathcal{A}_n
\end{align}
\end{subequations}
with
\begin{equation}
	\label{eq:vertex_sum}
	 \tilde{F}_{\nu_1 … \nu_s}^{wrq}
	 =  F_{\nu_1 … \nu_s}^{wrq}(k)- F_{\nu_1 … \nu_s}^{wrq}(0) \, .
\end{equation}
The argument of the vertex factors in~\eqref{eq:vertex_sum} indicates the dependence on $k$ in the vertex Feynman rule. The number of photons that attach to the off-shell heavy-fermion propagator in~\eqref{eq:ampgen2_diags} differs by one for the two diagrams. Following~\eqref{eq:eikid_gen} this results in the different sign of the two terms on the r.h.s.\ of~\eqref{eq:vertex_sum}.

The full all-order expression is then given by
\begin{equation}
	\label{eq:amp_allorder}
	\mathcal{A}^{i}_F
	=\sum_{u,t} \sum_{w=0}^u \sum_{r=0}^t \sum_{q=0}^s 
	\Big( \mathcal{A}_{wrq,F}^{ut(s-1),i}  + \tilde{\mathcal{A}}_{wrq,F}^{uts,i} \Big) \, ,
\end{equation}
where $s$ is fixed by the vertex type $F$ and the shifted index $(s-1)$ takes into account that the external photon attaches to $F$ in~\eqref{eq:diag_gen1}. The crucial property of the generalised off-shell eikonal identity~\eqref{eq:eikid_gen} is that it makes the simple dependence of the diagrams~\eqref{eq:diag_gen1} and~\eqref{eq:ampgen2_diags} on $w$ and $r$ evident. In particular,  the propagator structure $\mathcal{P}_{utsq,\mu_1…\mu_u}^{\nu_{1} … \nu_{q}}$ given in~\eqref{eq:prop_structure} does not depend at all on $w$ and $r$. Furthermore, we will only encounter the two cases where the vertex factors $F_{\nu_1 … \nu_s\nu}^{wrq}$ and $\tilde{F}_{\nu_1 … \nu_s}^{wrq}$ are either independent or linearly dependent on $w$ and $r$. This allows us to evaluate the all-order sums in~\eqref{eq:amp_allorder} with
\begin{subequations}
\label{eq:sums}
\begin{align}
	\label{eq:sums1}
	\sum_{x=0}^y S^{xy} &= \delta_{y,0}\, ,
	\\
	\label{eq:sums2}
	\sum_{x=0}^y S^{xy} x &= -\delta_{y,1} \, ,
\end{align}
\end{subequations}
where $\delta_{y,z}$ denotes the Kronecker delta function. Hence, these sum relations show that all virtual corrections beyond one loop cancel.

Let us apply this formalism to $\mathcal{A}_{n+1}^\text{mag}$ defined in~\eqref{eq:ampNLPmag}. The magnetic vertex~\eqref{eq:vertex_mag} fixes $s=1$ and the all-order amplitude~\eqref{eq:amp_allorder} reads
\begin{equation}
	\mathcal{A}_\text{mag}^{i}
	=  \sum_{u,t} \sum_{w=0}^u \sum_{r=0}^t
	\Big( \mathcal{A}_{wr0,\text{mag}}^{ut0,i} + \tilde{A}_{wr0,\text{mag}}^{ut1,i} + \tilde{A}_{wr1,\text{mag}}^{ut1,i} \Big) \, ,
\end{equation}
with
\begin{subequations}
\begin{align}
	F_\nu^{wr0} &= \frac{Q_i C_\text{mag}}{2m_i} \sigma_{\nu\mu} k^\mu \, ,
	\\
	\tilde{F}_{\nu_1}^{wrq} &= 0 \, .
\end{align}
\end{subequations}
Since these vertex factors depend neither on $w$ nor on $r$ we can apply~\eqref{eq:sums1} to show that all virtual corrections vanish. This yields the tree-level expression
\begin{equation}
	\label{eq:diag_mag}
    \mathcal{A}_{n+1}^\text{mag}
    = \sum_{i=1}^n \mathcal{A}_\text{mag}^{i}
    = \sum_{i=1}^n \begin{tikzpicture}[scale=.8,baseline={(0,-.1)}]
           
   \draw[line width=.3mm,style=double]  (-1.2,0) -- (1,0);
   \draw[line width=.3mm,style=double]  (1,0) -- (1.75,.75);
   \draw[line width=.3mm,style=double]  (1,0) -- (1.75,-.75);

   \draw[line width=.3mm]  [fill=black] (1.8,.25) circle (0.02);
   \draw[line width=.3mm]  [fill=black] (1.85,0.) circle (0.02);
   \draw[line width=.3mm]  [fill=black] (1.8,-.25) circle (0.02);

   \node at (-1.1,.3) {\footnotesize{$v_i$}};

   \draw[line width=0.3mm,photon]  (-.2,0) -- (-.2,1) node[right]{\footnotesize{$k$}};
   \draw (-.3,0) -- (-.2,0) 
     node[draw,isosceles triangle,fill=black,rotate=-90,minimum size=.3cm,inner sep = 0pt]{};

   \draw[line width=.3mm]  [fill=stuff] (1,0) circle (0.55) node[]{\footnotesize{$\mathcal{O}_\text{LP}$}};
       \end{tikzpicture}
\end{equation}
for the all-order amplitude.

The remaining NLP amplitude $\mathcal{A}_{n+1}^\text{kin}$ defined in~\eqref{eq:ampNLPkin} receives contributions from the three vertex types given in~\eqref{eq:vertex_kin}
\begin{equation}
	\mathcal{A}_{n+1}^\text{kin}
	= \mathcal{A}^{0\gamma}_{n+1}  
	+ \mathcal{A}^{1\gamma}_{n+1}
	+ \mathcal{A}^{2\gamma}_{n+1} \, .
\end{equation}
In the following we consider each contribution in turn.

In the case of the first vertex type with no attached photons ($s=0$), we only have diagrams of the form~\eqref{eq:ampgen2}. In particular, we find
\begin{equation}
	\mathcal{A}^{i}_{0\gamma}
	= \sum_{u,t} \sum_{w=0}^u \sum_{r=0}^t \tilde{\mathcal{A}}_{wr0,0\gamma}^{ut0} \, ,
\end{equation}
with
\begin{subequations}
	\label{eq:0gamma_vertex}
\begin{align}
	\Big(\frac{i}{2 m_i}\Big)^{-1} \tilde{F}^{wr0} 
	& =  \Big( \sum_{l=1}^w \ell_l + \sum_{l=1}^r \tilde{\ell}_l - k \Big)^2_{\perp_i}
	- \Big( \sum_{l=1}^w \ell_l + \sum_{l=1}^r \tilde{\ell}_l  \Big)^2_{\perp_i}
	\\
	\label{eq:vertex_sums}
	& = -2 \sum_{l=1}^w (\ell_l \cdot k - \ell_l \cdot v_i k \cdot v_i)
	- 2 \sum_{l=1}^r (\tilde{\ell}_l \cdot k - \tilde{\ell}_l \cdot v_i k \cdot v_i)
	- (k \cdot v_i)^2
	\\
	& \to  -2 w (\ell_1 \cdot k - \ell_1 \cdot v_i k \cdot v_i)
	- 2 r (\tilde{\ell}_1 \cdot k - \tilde{\ell}_1 \cdot v_i k \cdot v_i)
	- (k \cdot v_i)^2 \, .
\end{align}
\end{subequations}
Going from the second to the third line we have used that the vertex factor $\tilde{F}^{wr0}$ enters in the expression~\eqref{eq:ampgen2_expr} that is by definition completely symmetric under permutations of $\{\ell_1,…,\ell_w\}$ and $\{\tilde{\ell}_1,…,\tilde{\ell}_r\}$. As a consequence, each $\ell_l$ and $\tilde{\ell}_l$ term in~\eqref{eq:vertex_sums} yields the same result and we can make the replacements $\ell_l \to \ell_1$ and $\tilde{\ell}_l \to \tilde{\ell}_1$. As a result, the vertex factor only gives rise to a linear dependence on $w$ and $r$. Using~\eqref{eq:sums}, replacing $\ell _1 \to \ell$, $\tilde{\ell_1} \to \tilde{\ell}$, and plugging in~\eqref{eq:diag_rest} we obtain the one-loop exact result
\begin{align}
	\label{eq:kin0g_interm}
	\begin{split}
	\mathcal{A}^{i}_{0\gamma}
	=& \Bigg(
	 \frac{Q_i}{2m_i} \epsilon \cdot v_i
	\\
	& + \sum_{j \neq i} \frac{Q_i^2 Q_j}{m_i} \frac{\epsilon \cdot v_i}{k \cdot v_i}
	 \int[\mathrm{d}\ell] \frac{ v_i \cdot v_j( \ell \cdot k - \ell \cdot v_i k \cdot v_i)}
	{[\ell^2] [\ell \cdot v_i] [-\ell \cdot v_j] [\ell \cdot v_i - k \cdot v_i] }
	\\
	&+  \sum_{j \neq i} \frac{Q_i^3}{m_i} \frac{\epsilon \cdot v_i}{k \cdot v_i } \int[\mathrm{d}\tilde{\ell}] \frac{
	 \tilde{\ell} \cdot k - \tilde{\ell} \cdot v_i k \cdot v_i}
	{[\tilde{\ell}^2] [\tilde{\ell} \cdot v_i] [-\tilde{\ell} \cdot v_i] [\tilde{\ell} \cdot v_i - k \cdot v_i]}
	\Bigg)  \mathcal{A}_n \, .
	\end{split}
\end{align}
In the case of the loop momentum $\tilde{\ell}$ that only connects to the emitting direction $v_i$, the simple tensor decomposition
\begin{equation}
	\label{eq:pv}
	\tilde{\ell}^\rho \to \tilde{\ell} \cdot v_i v_i^\rho
\end{equation}
applies in the numerator. Plugging this into~\eqref{eq:kin0g_interm} for $\tilde{\ell} \cdot k$ directly leads to the cancellation of the corresponding contribution. As a consequence, the amplitude simplifies to
\begin{equation}
	\label{eq:a0g}
	\mathcal{A}^{i}_{0\gamma}
	= \Bigg( \frac{Q_i}{2m_i} \epsilon \cdot v_i 
	+\sum_{j \neq i} \frac{Q_i^2 Q_j}{m_i} \frac{\epsilon \cdot v_i}{(k \cdot v_i)^2} \int[\mathrm{d}\ell] \frac{ v_i \cdot v_j(\ell \cdot k - (k \cdot v_i)^2)}
	{ [\ell^2] [-\ell \cdot v_j] [\ell \cdot v_i - k \cdot v_i]}
	\Bigg)  \mathcal{A}_{n} \, .
\end{equation}
In~\eqref{eq:a0g} we have used the partial fraction identity~\eqref{eq:partial_fraction_simple} where the first term on the r.h.s.\ of this identity results in a scaleless contribution and can thus be neglected.

Next, we consider the second vertex type of~\eqref{eq:vertex_kin} that includes one photon, \emph{i.e.}\ $s=1$. In this case we obtain contributions both from~\eqref{eq:ampgen1} and~\eqref{eq:ampgen2} given by
\begin{equation}
	\mathcal{A}_{1\gamma}^{i}
	= \sum_{u,t} \sum_{w=0}^u \sum_{r=0}^t
	\Big( \mathcal{A}_{wr0,1\gamma}^{ut0,i} 
	+ \tilde{\mathcal{A}}_{wr0,1\gamma}^{ut1,i} 
	+ \tilde{\mathcal{A}}_{wr1,1\gamma}^{ut1,i} \Big) \, ,
\end{equation}
with
\begin{subequations}
\begin{align}
	\label{eq:1gamma_vertex}
	\Big(\frac{-iQ_i}{2m_i} \Big)^{-1} F^{wr0}_\nu
	& \to 2 w \ell_{1,\nu}^{\perp_i} + 2 r \tilde{\ell}_{1,\nu}^{\perp_i} - k_\nu^{\perp_i} \, ,
	\\
	\Big(\frac{-iQ_i}{2m_i} \Big)^{-1} \tilde{F}_{\nu_1}^{wrq} 
	&= - k_{\nu_1}^{\perp_i} \, .
\end{align}
\end{subequations}
The arrow in~\eqref{eq:1gamma_vertex} indicates that the linear dependence on $w$ and $r$ follows from the same reasoning as in~\eqref{eq:0gamma_vertex}. The corresponding one-loop exact result reads
\begin{equation}
        \label{eq:a1g}
	\mathcal{A}_{1\gamma}^{i}
	=  \Bigg(
	- \frac{Q_i}{2m_i} \epsilon \cdot v_i
	-  \sum_{j \neq i} \frac{Q_i^2 Q_j}{m_i} \frac{1}{k \cdot v_i } \int[\mathrm{d}\ell] \frac{v_i \cdot v_j (\epsilon \cdot \ell - 2 \epsilon \cdot v_i k \cdot v_i) + \epsilon \cdot v_i k \cdot v_j}
	{[\ell^2] [-\ell \cdot v_j] [\ell \cdot v_i - k \cdot v_i]}
	\Bigg) \mathcal{A}_n\, ,
\end{equation}
where we have used partial fraction decomposition~\eqref{eq:partial_fraction_simple} and the property $\epsilon \cdot k = 0$. Furthermore, we have replaced in the numerator
\begin{equation}
	\label{eq:numsimp}
	\ell \cdot v_i = [\ell \cdot v_i - k \cdot v_i] + k \cdot v_i
\end{equation}
and neglected the first term on the r.h.s.\ since it results in a scaleless contribution after cancelling the corresponding propagator. Note that the $\tilde{\ell}$ terms also vanish in this case. This follows from the tensor decomposition~\eqref{eq:pv} and the property $ k_{\perp_i} \cdot v_i = 0$.

Finally, the third vertex type of~\eqref{eq:vertex_kin} that connects two photons ($s=2$) gives rise to the contribution
\begin{equation}
	\mathcal{A}_{2\gamma}^{i}
	= \sum_{u,t} \sum_{w=0}^u \sum_{r=0}^t
	\Big( \mathcal{A}_{wr0,2\gamma}^{ut1,i} + \mathcal{A}_{wr1,2\gamma}^{ut1,i}
	+ \tilde{\mathcal{A}}_{wr0,2\gamma}^{ut2,i} + \tilde{\mathcal{A}}_{wr1,2\gamma}^{ut2,i} + \tilde{\mathcal{A}}_{wr2,2\gamma}^{ut2,i} \Big) \, ,
\end{equation}
with
\begin{subequations}
\begin{align}
	F^{wrq}_{\nu_1,\nu} & = \frac{i Q_i^2}{m_i} g_{\nu_1\nu}^{\perp_i} \, ,
	\\
	\tilde{F}_{\nu_1 \nu_2}^{wrq} &= 0 \, .
\end{align}
\end{subequations}
This directly yields the one-loop exact result
\begin{equation}
	\label{eq:a2g}
	\mathcal{A}_{2\gamma}^{i}
	= \sum_{j \neq i}  \frac{Q_i^2 Q_j}{m_i} \int[\mathrm{d}\ell]  \frac{\epsilon \cdot v_j - v_i \cdot v_j \epsilon \cdot v_i }
	{[\ell^2] [-\ell \cdot v_j] [\ell \cdot v_i - k \cdot v_i]}
	\mathcal{A}_n \, ,
\end{equation}
where contributions corresponding to $\tilde{\ell}$ have vanished due to $\epsilon_{\perp_i} \cdot v_i = 0$.

While~\eqref{eq:a2g} only contributes at one loop, the tree-level terms in~\eqref{eq:a0g} and~\eqref{eq:a1g} exactly cancel. The sum of the three contributions therefore only enters at one loop. To simplify the corresponding expression we apply the tensor decomposition
\begin{equation}
	\ell^\rho \to
	\frac{\ell \cdot v_j v_i \cdot v_j - \ell \cdot v_i}{(v_i \cdot v_j)^2-1} v_i^\rho
	+\frac{\ell \cdot v_i v_i \cdot v_j - \ell \cdot v_j}{(v_i \cdot v_j)^2-1} v_j^\rho
\end{equation}
to $\ell \cdot \epsilon$ and $\ell \cdot k$ in~\eqref{eq:a0g} and~\eqref{eq:a1g}, use~\eqref{eq:numsimp} to replace $\ell \cdot v_i$ with $k \cdot v_i$,  and substitute $v_i = p_i/m_i + \mathcal{O}(\lambda)$. This yields
\begin{equation}
	\label{eq:soft_res}
	\mathcal{A}_{\text{kin}}^{i}
	=   \sum_{j \neq i} Q_i^2 Q_j  A_n \Big(\frac{\epsilon \cdot p_i}{k \cdot p_i} - \frac{\epsilon \cdot p_j}{k \cdot p_j} \Big)
	    S^{(1)}(p_i,p_j,k) \, ,
\end{equation}
where the one-loop exact soft function
\begin{equation}
	\label{eq:soft_func}
	S^{(1)}(p_i,p_j,k) = \frac{m_i^2 k \cdot p_j}{\big( (p_i \cdot p_j)^2-m_i^2 m_j^2 \big) k \cdot p_i}
	\Big( p_i \cdot p_j I_1(p_i,k) + m_j^2 k \cdot p_i I_2(p_i,p_j,k) \Big)
\end{equation}
is defined in terms of the two simple integrals
\begin{subequations}
\label{eq:integrals}
\begin{align}
	I_1(p_i,k) 
	&= \int [\mathrm{d} \ell]
	\frac{1}{[\ell^2+i0][\ell \cdot p_i - k \cdot p_i + i 0]} \, ,
	\\
	I_2(p_i,p_j,k) 
	&= \int [\mathrm{d} \ell]
	\frac{1}{[\ell^2+i0][-\ell \cdot p_j+i 0][\ell \cdot p_i - k \cdot p_i + i 0]} \, .
\end{align}
\end{subequations}
The causal $+i0$ prescription is displayed explicitly in the above integrals. This exactly agrees with the one-loop soft function given in~(3.29) of~\cite{Engel:2021ccn}. In that case the soft contribution was extracted in full QED using the method of regions. The above calculation proves that this result is one-loop exact. The analytic results for the integrals~\eqref{eq:integrals} with exact $\epsilon$ dependence can be found in Appendix~A of~\cite{Engel:2021ccn}.

The complete NLP soft contribution to the radiative scattering amplitude $\mathcal{A}_{n+1}$ at all orders is given by
\begin{equation}
	\label{eq:total_soft}
	\mathcal{A}_{n+1}^\text{soft}
	= \mathcal{A}_{n+1}^\text{kin} 
	= \sum_{i=1}^n \mathcal{A}_{\text{kin}}^{i} \, .
\end{equation}
The corresponding expression for the squared amplitude is obtained by interfering this result with the LP eikonal approximation~\eqref{eq:diag_lp}. The explicit formula is given below in~\eqref{eq:lbk_soft}. We now turn to the evaluation of the hard matching corrections. 

\section{The hard contribution}
\label{sec:hard}

The previous section has demonstrated that only $\mathcal{A}_{n+1}^\text{kin}$ yields non-vanishing soft virtual corrections that enter the total soft amplitude $\mathcal{A}_{n+1}^\text{soft}$. These corrections turn out to be one-loop exact. Hence, $\mathcal{A}_{n+1}^\text{soft}$ exactly agrees with the one-loop soft contribution from~\cite{Engel:2021ccn} obtained in full QED using the method of regions. In particular, QED diagrams with only hard loop momenta do not contribute to $\mathcal{A}_{n+1}^\text{kin}$. This is a consequence of the cancellation of the tree-level contributions in~\eqref{eq:a0g} and~\eqref{eq:a1g}. Hence, $\mathcal{A}_{n+1}^\text{kin}$ contains exactly one soft loop. The remaining amplitudes $\mathcal{A}_{n+1}^\text{LP}$, $\mathcal{A}_{n+1}^\text{NLP}$, and $\mathcal{A}_{n+1}^\text{mag}$, defined in~\eqref{eq:ampLP},~\eqref{eq:ampNLPO}, and~\eqref{eq:ampNLPmag}, thus combine to the purely hard momentum region at NLP in QED. This follows from the absence of any soft virtual corrections in~\eqref{eq:diag_lp}, \eqref{eq:diag_nlp}, and \eqref{eq:diag_mag}. As a result, the total hard contribution to NLP is given by
\begin{equation}
	\label{eq:hard_total}
	\mathcal{A}_{n+1}^\text{hard}
	= \Bigg( \sum_{i=1}^n \begin{tikzpicture}[scale=.8,baseline={(0,-.1)}]
        	      
   \draw[line width=.3mm]  (-1.2,0) -- (1,0);
   \draw[line width=.3mm]  (1,0) -- (1.75,.75);
   \draw[line width=.3mm]  (1,0) -- (1.75,-.75);

   \draw[line width=.3mm]  [fill=black] (1.8,.25) circle (0.02);
   \draw[line width=.3mm]  [fill=black] (1.85,0.) circle (0.02);
   \draw[line width=.3mm]  [fill=black] (1.8,-.25) circle (0.02);

   \node at (-1.1,.3) {\footnotesize{$p_i$}};

   \draw[line width=0.3mm,photon]  (-.2,0) -- (-.2,1) node[right]{\footnotesize{$k$}};

   \draw[line width=.3mm]  [fill=stuff] (-.2,0) circle (0.3) node[]{\footnotesize{$\Lambda$}};
   \draw[line width=.3mm]  [fill=stuff] (1,0) circle (0.55) node[]{\footnotesize{$\Gamma_\text{ext}$}};
       	   \end{tikzpicture}
	   \, \, \Bigg)
	   +
	   \begin{tikzpicture}[scale=.8,baseline={(0,-.1)}]
        	      
   \draw[line width=.3mm]  (-.25,0) -- (1,0);
   \draw[line width=.3mm]  (1,0) -- (1.75,.75);
   \draw[line width=.3mm]  (1,0) -- (1.75,-.75);

   \draw[line width=.3mm]  [fill=black] (1.8,.25) circle (0.02);
   \draw[line width=.3mm]  [fill=black] (1.85,0.) circle (0.02);
   \draw[line width=.3mm]  [fill=black] (1.8,-.25) circle (0.02);

   \draw[line width=0.3mm,photon]  (1,0) -- (1,1) node[right]{\footnotesize{$k$}};

   \draw[line width=.3mm]  [fill=stuff] (1,0) circle (0.55) node[]{\footnotesize{$\Gamma_\text{int}$}};
       	   \end{tikzpicture}
	   = \mathcal{A}_{n+1}^\text{LP} + \mathcal{A}_{n+1}^\text{NLP} + \mathcal{A}_{n+1}^\text{mag}
		+ \mathcal{O}(\lambda) \, ,
\end{equation}
where the gray blobs contain hard loop corrections only and $\Lambda$ denotes the vertex form factor. The complete hard contribution can therefore be calculated by expanding the corresponding QED diagrams at the loop integrand level in the soft scale assuming a hard scaling for all loop momenta $\ell_l \sim p_j,m_j$. In the following, we disentangle the magnetic contribution $\mathcal{A}_{n+1}^\text{mag}$  from the rest and show that it vanishes at the level of the unpolarised squared amplitude. As we will see, the remaining contributions are related to the non-radiative process by virtue of the LBK theorem.

The magnetic Wilson coefficient $C_\text{mag}$ can be fixed by considering the vertex with one off-shell fermion. The corresponding matching equation reads
\begin{equation}
	\begin{tikzpicture}[scale=.8,baseline={(0,-.1)}]

   \draw[line width=.3mm]  (-1,-1) node[left] {\footnotesize{$p$}} -- (0,0);
   \draw[line width=.3mm]  (1,-1) node[right] {\footnotesize{$p + k$}} -- (0,0);
   \draw[line width=.3mm, photon] (0,0) -- (0,1) node[right] {\footnotesize{$k$}};
   \draw[line width=.3mm]  [fill=stuff] (0,0) circle (0.3) node[]{\footnotesize{$\Lambda$}};

	 \end{tikzpicture}
	 \stackrel{!}{=}
	 \begin{tikzpicture}[scale=.8,baseline={(0,-.1)}]

   \draw[line width=.3mm,style=double]  (-1,-1)  -- (0,0);
   \draw[line width=.3mm,style=double]  (1,-1)  -- (0,0);
   \draw[line width=.3mm, photon] (0,0) -- (0,1);

	 \end{tikzpicture}
	 +
	\begin{tikzpicture}[scale=.8,baseline={(0,-.1)}]

   \draw[line width=.3mm,style=double]  (-1,-1)  -- (0,0);
   \draw[line width=.3mm,style=double]  (1,-1)  -- (0,0);
   \node[rectangle,fill=black] at (0,0) {};
   \draw[line width=.3mm, photon] (0,0) -- (0,1);

	 \end{tikzpicture}
	 +
	\begin{tikzpicture}[scale=.8,baseline={(0,-.1)}]

   \draw[line width=.3mm,style=double]  (-1,-1) -- (0,0);
   \draw[line width=.3mm,style=double]  (1,-1) -- (0,0);
   \node[rectangle,fill=black,rotate=45] at (.5,-.5) {};
   \draw[line width=.3mm, photon] (0,0) -- (0,1);

	 \end{tikzpicture}
	 +
	\begin{tikzpicture}[scale=.8,baseline={(0,-.1)}]

   \draw[line width=.3mm,style=double]  (-1,-1) -- (0,0);
   \draw[line width=.3mm,style=double]  (1,-1) -- (0,0);
   \node[draw,isosceles triangle,fill=black,rotate=-90,minimum size=.3cm,inner sep = 0pt]{};
   \draw[line width=.3mm, photon] (0,0) -- (0,1);

	 \end{tikzpicture}
	 \,\, + \,\, \mathcal{O}(\lambda^2)
\end{equation}
with $p^2=m^2$ and $k^2=0$. The first diagram on the r.h.s.\ coming from the LP Lagrangian $\mathcal{L}_\text{LP}$ does not contribute beyond tree level. This is also the case for the second and third diagram generated by $\mathcal{L}_\text{kin}$ as a consequence of reparametrisation invariance (\emph{c.f.}\ Section~\ref{sec:eft}). Separating the tree-level contribution by writing the form factor $\Lambda =  \gamma^\nu + \tilde{\Lambda}$ and $C_\text{mag}=1+\tilde{C}_\text{mag}$, we find the simple relation
\begin{equation}
	\begin{tikzpicture}[scale=.8,baseline={(0,-.1)}]

   \draw[line width=.3mm]  (-1,-1) node[left] {\footnotesize{$p$}} -- (0,0);
   \draw[line width=.3mm]  (1,-1) node[right] {\footnotesize{$p + k$}} -- (0,0);
   \draw[line width=.3mm, photon] (0,0) -- (0,1) node[right] {\footnotesize{$k$}};
   \draw[line width=.3mm]  [fill=stuff] (0,0) circle (0.3) node[]{\footnotesize{$\tilde{\Lambda}$}};

	 \end{tikzpicture}
	 \stackrel{!}{=}
	\begin{tikzpicture}[scale=.8,baseline={(0,-.1)}]

   \draw[line width=.3mm,style=double]  (-1,-1) -- (0,0);
   \draw[line width=.3mm,style=double]  (1,-1) -- (0,0);
   \node[draw,isosceles triangle,fill=black,rotate=-90,minimum size=.3cm,inner sep = 0pt]{};
   \draw[line width=.3mm, photon] (0,0) -- (0,1);
   \node at (.6,.4) {\footnotesize{$\tilde{C}_\text{mag}$}};

	 \end{tikzpicture}
	 \,\,+ \,\, \mathcal{O}(\lambda^2)
	 \,\,\,\, ,
\end{equation}
which implies
\begin{equation}
	\label{eq:diag_nonfac}
	\begin{tikzpicture}[scale=.8,baseline={(0,-.1)}]
        	   
   \draw[line width=.3mm]  (-1.2,0) -- (1,0);
   \draw[line width=.3mm]  (1,0) -- (1.75,.75);
   \draw[line width=.3mm]  (1,0) -- (1.75,-.75);

   \draw[line width=.3mm]  [fill=black] (1.8,.25) circle (0.02);
   \draw[line width=.3mm]  [fill=black] (1.85,0.) circle (0.02);
   \draw[line width=.3mm]  [fill=black] (1.8,-.25) circle (0.02);

   \node at (-1.1,.3) {\footnotesize{$p_i$}};

   \draw[line width=0.3mm,photon]  (-.2,0) -- (-.2,1) node[right]{\footnotesize{$k$}};

   \draw[line width=.3mm]  [fill=stuff] (-.2,0) circle (0.3) node[]{\footnotesize{$\tilde{\Lambda}$}};
   \draw[line width=.3mm]  [fill=stuff] (1,0) circle (0.55) node[]{\footnotesize{$\Gamma_\text{ext}$}};
       	\end{tikzpicture}\,
	=
	\begin{tikzpicture}[scale=.8,baseline={(0,-.1)}]
       	   
   \draw[line width=.3mm,style=double]  (-1.2,0) -- (1,0);
   \draw[line width=.3mm,style=double]  (1,0) -- (1.75,.75);
   \draw[line width=.3mm,style=double]  (1,0) -- (1.75,-.75);

   \draw[line width=.3mm]  [fill=black] (1.8,.25) circle (0.02);
   \draw[line width=.3mm]  [fill=black] (1.85,0.) circle (0.02);
   \draw[line width=.3mm]  [fill=black] (1.8,-.25) circle (0.02);

   \node at (-1.1,.3) {\footnotesize{$v_i$}};

   \draw[line width=0.3mm,photon]  (-.2,0) -- (-.2,1) node[right]{\footnotesize{$k$}};
   \node[draw,isosceles triangle,fill=black,rotate=-90,minimum size=.3cm,inner sep = 0pt] at (-.2,0) {};

   \draw[line width=.3mm]  [fill=stuff] (1,0) circle (0.55) node[]{\footnotesize{$\mathcal{O}_\text{LP}$}};
   \node at (-.4,-.4) {\footnotesize{$\tilde{C}_\text{mag}$}};
      	\end{tikzpicture}
	\,\, +  \,\,\mathcal{O}(\lambda)
\end{equation}
for the full process. The tree-level EFT diagram on the r.h.s.\ evaluates straightforwardly to
\begin{equation}
	\label{eq:magnetic_eft}
	\begin{tikzpicture}[scale=.8,baseline={(0,-.1)}]
       	   
   \draw[line width=.3mm,style=double]  (-1.2,0) -- (1,0);
   \draw[line width=.3mm,style=double]  (1,0) -- (1.75,.75);
   \draw[line width=.3mm,style=double]  (1,0) -- (1.75,-.75);

   \draw[line width=.3mm]  [fill=black] (1.8,.25) circle (0.02);
   \draw[line width=.3mm]  [fill=black] (1.85,0.) circle (0.02);
   \draw[line width=.3mm]  [fill=black] (1.8,-.25) circle (0.02);

   \node at (-1.1,.3) {\footnotesize{$v_i$}};

   \draw[line width=0.3mm,photon]  (-.2,0) -- (-.2,1) node[right]{\footnotesize{$k$}};
   \node[draw,isosceles triangle,fill=black,rotate=-90,minimum size=.3cm,inner sep = 0pt] at (-.2,0) {};

   \draw[line width=.3mm]  [fill=stuff] (1,0) circle (0.55) node[]{\footnotesize{$\mathcal{O}_\text{LP}$}};
   \node at (-.4,-.4) {\footnotesize{$\tilde{C}_\text{mag}$}};
      	\end{tikzpicture} \,
	=
	\frac{Q_i \tilde{C}_\text{mag}}{2m_i} \Gamma_\text{ext} \epsilon \cdot H u_{v_i}
\end{equation}
with
\begin{equation}
	\label{eq:htensor}
	H^\mu 
	= \gamma^\mu 
	- \frac{1}{k \cdot v_i}  \slashed{k}v_i^\mu
	- \frac{1}{k \cdot v_i} \gamma^\mu \slashed{k} \, .
\end{equation}
At the level of the unpolarised squared amplitude this contribution enters via interference with the LP eikonal approximation~\eqref{eq:diag_lp} yielding
\begin{equation}
	\label{eq:eikinterference}
	\mathcal{M}_{n+1}^\text{mag}
	 =  \sum_{i,j=1}^n \frac{Q_i \tilde{C}_\text{mag}}{2m_i}
	 \frac{Q_j \epsilon^\ast \cdot v_j}{-k \cdot v_j} \Gamma_\text{ext} \epsilon \cdot H (1+\slashed{v_i}) \Gamma_\text{ext}^\dagger
	 + h.c. + \mathcal{O}(\lambda^0) \, .
\end{equation}
The specific structure of $H^\mu$ in~\eqref{eq:htensor} results in a cancellation of the term on the r.h.s.\ of~\eqref{eq:eikinterference} with its hermitian conjugate. This follows from basic Dirac algebra as shown in Section 5.2.1 of~\cite{Engel:2022kde} in the context of the one-loop LBK theorem. As a result, the magnetic contribution vanishes to all orders in the case of unpolarised scattering. We emphasize, however, that even for polarised fermions the formalism developed here is useful. This is in particular the case since the structure of the magnetic contribution~\eqref{eq:magnetic_eft} is simple and the corresponding Wilson coefficient $C_\text{mag}$ is known to three loops~\cite{Grozin:2007fh}.

As a check, we have explicitly verified up to two loops that the full QED diagram on the l.h.s.\ of~\eqref{eq:diag_nonfac} yields the same structure as~\eqref{eq:magnetic_eft}. To perform this calculation, we have employed \texttt{Package-X}~\cite{Patel:2015tea} for the Dirac algebra, FIRE~\cite{Smirnov:2019qkx} for the IBP reduction, and \texttt{HyperInt}~\cite{Panzer:2014caa} to evaluate the scalar integrals. Matching this result to the EFT contribution, we obtain
\begin{equation}
	\tilde{C}_\text{mag} 
	= \frac{\alpha}{4\pi}(2)
	+ \Big( \frac{\alpha}{4\pi} \Big)^2
	\Big( -31 +\frac{20}{3}\pi^2-8 \pi^2 \log 2+12 \zeta_3 \Big)
	+ \mathcal{O}(\alpha^3)
\end{equation}
in agreement with~\cite{Czarnecki:1997dz}.

At the level of the unpolarised squared amplitude the relation between the hard region in the full theory and the EFT thus reduces to
\begin{equation}
	\label{eq:hard_unpol}
	\mathcal{M}_{n+1}^\text{hard}
	= \sum_\text{pol}\Bigg| \,
	  \Bigg( \sum_{i=1}^n \begin{tikzpicture}[scale=.8,baseline={(0,-.1)}]
        	      
   \draw[line width=.3mm]  (-1.2,0) -- (1,0);
   \draw[line width=.3mm]  (1,0) -- (1.75,.75);
   \draw[line width=.3mm]  (1,0) -- (1.75,-.75);

   \draw[line width=.3mm]  [fill=black] (1.8,.25) circle (0.02);
   \draw[line width=.3mm]  [fill=black] (1.85,0.) circle (0.02);
   \draw[line width=.3mm]  [fill=black] (1.8,-.25) circle (0.02);

   \node at (-1.1,.3) {\footnotesize{$p_i$}};

   \draw[line width=0.3mm,photon]  (-.2,0) -- (-.2,1) node[right]{\footnotesize{$k$}};

   \draw[line width=.3mm]  [fill=stuff] (1,0) circle (0.55) node[]{\footnotesize{$\Gamma_\text{ext}$}};
       	   \end{tikzpicture}
	   \, \, \Bigg)
	   +
	   \begin{tikzpicture}[scale=.8,baseline={(0,-.1)}]
        	      
   \draw[line width=.3mm]  (-.25,0) -- (1,0);
   \draw[line width=.3mm]  (1,0) -- (1.75,.75);
   \draw[line width=.3mm]  (1,0) -- (1.75,-.75);

   \draw[line width=.3mm]  [fill=black] (1.8,.25) circle (0.02);
   \draw[line width=.3mm]  [fill=black] (1.85,0.) circle (0.02);
   \draw[line width=.3mm]  [fill=black] (1.8,-.25) circle (0.02);

   \draw[line width=0.3mm,photon]  (1,0) -- (1,1) node[right]{\footnotesize{$k$}};

   \draw[line width=.3mm]  [fill=stuff] (1,0) circle (0.55) node[]{\footnotesize{$\Gamma_\text{int}$}};
       	   \end{tikzpicture}
	   \, \Bigg|^2
	   = \sum_\text{pol} | \mathcal{A}_{n+1}^\text{LP}+ \mathcal{A}_{n+1}^\text{NLP} |^2
	   + \mathcal{O}(\lambda^0)
	    \, .
\end{equation}
It is however exactly for these types of diagrams that the original tree-level proof of the LBK theorem
presented in Section~\ref{sec:lbk_tree} still applies. This is in particular the case because $\Gamma_\text{int}$ by definition only contains hard loop corrections  and therefore cannot give rise to $1/k$ poles. In complete analogy to the tree-level case~\eqref{eq:split_gaugeinv}, we can fix the hard internal emission contribution via gauge invariance. Consequently, the hard contribution~\eqref{eq:hard_unpol} is simply given by the r.h.s.\ of~\eqref{eq:lbk_tree} with $\mathcal{M}_{n}^{(0)} \to \mathcal{M}_{n}$. The explicit formula is given in the following section in~\eqref{eq:lbk_hard} where the complete all-order LBK theorem is presented.

\section{The LBK theorem to all orders}
\label{sec:lbk}

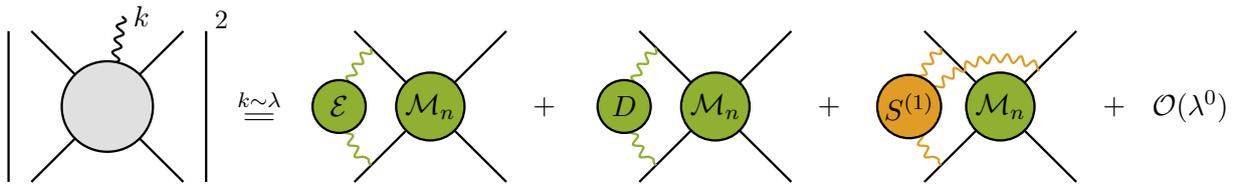
\begin{figure}
    \label{fig:lbk_allorder}
    \centering
    \begin{tikzpicture}[scale=1,baseline={(1,0)}]
    	    
    \draw[line width=.3mm] (-1.3,-1) -- (-1.3,1);
    \draw[line width=.3mm] (1.3,-1) -- (1.3,1);
    \node at (1.5,1.15) {\small{$2$}};
    \draw[line width=.3mm] (-1,-1)-- (0,0);
    \draw[line width=.3mm] (0,0) -- (1,-1);
    \draw[line width=.3mm] (-1,+1) -- (0,0);
    \draw[line width=.3mm] (0,0) -- (1,+1);
    \draw[line width=.3mm]  [tightphoton] (.2,+1.2) node[right]{$k$} -- (0,0);
    
    \draw[line width=.3mm]  [fill=stuff] (0,0) circle (0.6);
    
    \node at (2,0) {$\wideeq{k\sim \lambda}$};

    \draw[line width=.3mm] (-1+4.25,-1)-- (0+4.25,0);
    \draw[line width=.3mm] (0+4.25,0) -- (1+4.25,-1);
    \draw[line width=.3mm] (-1+4.25,+1) -- (0+4.25,0);
    \draw[line width=.3mm] (0+4.25,0) -- (1+4.25,+1);
    
    \centerarc [line width=0.3mm,tightphoton, hard](0+4.25,0)(135:225:1.1);
    \draw[line width=.3mm]  [fill=hard] (-1.2+4.25,0) circle (0.35) node[]{$\eik$};
    \draw[line width=.3mm]  [fill=hard] (0+4.25,0) circle (0.45) node[]{$\mathcal{M}_n$};
    
    \node at (5.75,0) {$+$};
    
    \draw[line width=.3mm] (-1+8.,-1)-- (0+8.,0);
    \draw[line width=.3mm] (0+8.,0) -- (1+8.,-1);
    \draw[line width=.3mm] (-1+8.,+1) -- (0+8.,0);
    \draw[line width=.3mm] (0+8.,0) -- (1+8.,+1);
    \centerarc [line width=0.3mm,tightphoton, hard](0+8.,0)(135:225:1.1);
    \draw[line width=.3mm]  [fill=hard] (-1.2+8.,0) circle (0.35) node[]{$D$};
    \draw[line width=.3mm]  [fill=hard] (0+8.,0) circle (0.45) node[]{$\mathcal{M}_n$};
    
    \node at (9.5,0) {$+$};
    
    \draw[line width=.3mm] (-1+11.75,-1)-- (0+11.75,0);
    \draw[line width=.3mm] (0+11.75,0) -- (1+11.75,-1);
    \draw[line width=.3mm] (-1+11.75,+1) -- (0+11.75,0);
    \draw[line width=.3mm] (0+11.75,0) -- (1+11.75,+1);
    \centerarc [line width=0.3mm,tightphoton, soft](0+11.75,0)(135:225:1.1);
    \centerarc [line width=0.3mm,tightphoton, soft](11.75,-.5)(160:62:1.1);
    \draw[line width=.3mm]  [fill=hard] (0+11.75,0) circle (0.45) node[]{$\mathcal{M}_n$};
    \draw[line width=.3mm]  [fill=soft] (-1.2+11.75,0) circle (0.42) node[]{$S^{(1)}$};

    \node at (13.25,0) {$+$};
    \node at (14.25,0) {$\mathcal{O}(\lambda^0)$};
    \end{tikzpicture}
    \caption{Schematic illustration of the all-order LBK theorem~\eqref{eq:lbk_allorder}. The soft function $S^{(1)}$ is one-loop exact.}
\end{figure}

In the previous two sections we have proven the following two statements about one-photon radiation at NLP in the soft limit. First, all virtual soft corrections beyond one loop vanish. The remaining one-loop result is given by a universal soft function multiplied by the non-radiative (all-order) amplitude. Second, all hard matching corrections can be derived from the non-radiative process by analogy to the tree-level LBK theorem. This latter statement only holds at the level of the unpolarised squared amplitude $\mathcal{M}_{n+1} = \sum_\text{pol} |\mathcal{A}_{n+1}|^2$.

These statements allow us to generalise the LBK theorem in an all-order closed form. In fact, the generalisation is given by the one-loop formula~(3.31) of~\cite{Engel:2021ccn} by simply dropping the loop order indices, \emph{e.g.}\ replacing $\mathcal{M}_{n+1}^{(1)} \to \mathcal{M}_{n+1}$. To all orders, the soft limit up to NLP is thus given by\footnote{The book-keeping parameter $\lambda$ is only used to make the power counting more transparent and is to be set to one in calculations.}
\begin{subequations}
\label{eq:lbk_allorder}
\begin{equation}
    \mathcal{M}_{n+1}(\{p\},k)
    \wideeq{k\sim\lambda} 
    \mathcal{M}_{n+1}^{\text{hard}} + \mathcal{M}_{n+1}^{\text{soft}} + \mathcal{O}(\lambda^0)
\end{equation}
with
\begin{align}
	\label{eq:lbk_hard}
	\mathcal{M}_{n+1}^{\text{hard}}
	&= \sum_{l,i} Q_i Q_l \Big( 
	 - \frac{1}{\lambda^2} \frac{p_i \cdot p_l}{(k \cdot p_i) (k \cdot p_l)}
	 + \frac{1}{\lambda} \frac{p_l \cdot \tilde{D}_i} {k \cdot p_l}
	 \Big) \mathcal{M}_n(\{s\},\{m^2\}) \, ,
	 \\
	 \label{eq:lbk_soft}
	 \mathcal{M}_{n+1}^{\text{soft}}
	 &= \frac{1}{\lambda} \sum_{l,i,j \neq i} Q_i^2 Q_j Q_l \Big(
	 \frac{p_i \cdot p_l}{(k \cdot p_i) ( k \cdot p_l)}
	 - \frac{p_j \cdot p_l}{(k \cdot p_j ) (k \cdot p_l)}
	 \Big) 2 S^{(1)}(p_i,p_j,k) \mathcal{M}_n(\{s\},\{m^2\}) \, .
\end{align}
\end{subequations}
The LBK differential operator $\tilde{D_i}$ expressed in terms of kinematic invariants is given in~\eqref{eq:lbkop_new} and the one-loop exact soft function $S^{(1)}(p_i,p_j,k)$ is defined in~\eqref{eq:soft_func}. All fermions are assumed to be incoming. For outgoing particles one can simply replace the corresponding momentum $p_i$ with $-p_i$. At tree level, \eqref{eq:lbk_allorder} reduces to the original version of the LBK theorem~\eqref{eq:lbk_tree} since the soft contribution only enters at one loop. We again emphasize that the invariants $\{s\}=\{s(\{p\},\{m^2\})\}$ are only uniquely defined up to momentum violating $\mathcal{O}(\lambda)$ terms since $\sum_i p_i = k$. While this is irrelevant for the soft contribution that starts at $\mathcal{O}(1/\lambda)$, it matters for the hard part. For this reason, it is crucial to ensure that the same definition is used both in the evaluation of the non-radiative contribution as well as in the calculation of the derivatives $\partial s_L/\partial p_i^\mu$ in~\eqref{eq:lbkop_new}. Figure~\ref{fig:lbk_allorder} shows a conceptual illustration of the all-order LBK theorem~\eqref{eq:lbk_allorder}. The hard contribution~\eqref{eq:lbk_hard} is represented by the first two diagrams on the r.h.s. The soft contribution~\eqref{eq:lbk_soft}, connecting three external legs simultaneously, is depicted in the third diagram.

\section{Real-virtual-virtual corrections for muon-electron scattering}
\label{sec:muone}

Muon-electron scattering has gained considerable attention~\cite{Banerjee:2020rww,Broggio:2022htr,CarloniCalame:2020yoz,Alacevich:2018vez,Budassi:2021twh,Budassi:2022kqs,Fael:2019nsf,Fael:2018dmz,Masiero:2020vxk,Dev:2020drf,Schubert:2019nwm,GrillidiCortona:2022kbq,Galon:2022xcl,Asai:2021wzx} in recent years due to the MUonE experiment~\cite{CarloniCalame:2015obs,Abbiendi:2016xup,Spedicato:2022Nb,Abbiendi:2022oks} requiring a high-precision theory prediction at the level of 10 parts per million (ppm)~\cite{Banerjee:2020tdt}. Two independent Monte Carlo codes~\cite{Banerjee:2020rww,CarloniCalame:2020yoz} have been developed with the aim of ensuring this level of precision. Currently, both codes include QED corrections up to NNLO. In~\cite{CarloniCalame:2020yoz} the genuine two-loop four-point topologies have been taken into account using a YFS-inspired approximation. The calculation presented in~\cite{Broggio:2022htr}, on the other hand, takes these into account by massifying~\cite{Penin:2005eh,Mitov:2006xs,Becher:2007cu,Engel:2018fsb} the amplitude with a vanishing electron mass~\cite{Mastrolia:2017pfy,DiVita:2018nnh,Bonciani:2021okt,Mandal:2022vju}. This gives a correct description up to terms that are polynomially suppressed by the electron mass. 

These results show sizable NNLO corrections due to soft and collinear enhancements.
Contributions at even higher orders have to be taken into account to reach the 10 ppm target precision. This includes the resummation of the corresponding leading soft and collinear logarithms with a parton shower. Furthermore, a collaborative effort to calculate the dominant electron-line corrections at N$^3$LO has been launched recently~\cite{Durham:n3lo}. Two major steps in this direction have already been taken. A subtraction scheme for soft singularities in QED has been formulated to all orders~\cite{Engel:2019nfw}. Furthermore, the heavy-quark form factor has been calculated at three loops~\cite{Fael:2022miw,Fael:2022rgm} with a semi-numerical approach. The only missing piece is the real-virtual-virtual amplitude, which is currently only known for massless fermions~\cite{Garland:2001tf,Garland:2002ak}.

As a first application of the generalised LBK theorem we therefore consider the two-loop electron-line correction to the process
\begin{equation}
	e^-(p_1) \mu^-(p_2) \to e^-(p_3) \mu^-(p_4) \gamma(k)
\end{equation}
and calculate the corresponding amplitude at NLP in the soft limit for $k$. We take~\eqref{eq:lbk_allorder} with the charge and momentum signs as
\begin{subequations}
\label{eq:signs}
\begin{alignat}{10}
&p_1&\to&+p_1,&\qquad   &p_2&\to&+p_2,&\qquad
&p_3&\to&-p_3,&\qquad   &p_4&\to&-p_4\, ,&\qquad
\\
&Q_1&=  &- e,&   &Q_2&=  &- e,&
&Q_3&=  &+ e,&   &Q_4&=  &+ e \, .
\end{alignat}
\end{subequations}
These sign conventions also have to be taken into account in the derivatives $\partial/\partial p_i^\mu$ of~\eqref{eq:lbkop_new}. Furthermore, we define the invariants as
\begin{align}
	\label{eq:invariants}
    \{s\}=\{s=(p_1+p_2)^2,\,
            t=(p_2-p_4)^2 \} \, .
\end{align}
This definition has to be consistently used both in the evaluation of the non-radiative amplitude as well as in the calculation of the derivatives $\partial s_L/\partial p_i^\mu$. The LBK operators~\eqref{eq:lbkop_new} for the incoming and outgoing electron then read
\begin{subequations}
\begin{align}
	\tilde{D}_1^\mu
	&= \Big(\frac{p_1^\mu}{k \cdot p_1} k \cdot p_2 - p_2^\mu \Big) 2 \frac{\partial}{\partial s}\, ,
	\\
	\tilde{D}_3^\mu
	&= 0 \, .
\end{align}
\end{subequations}
The operators $\tilde{D}_2^\mu$ and $\tilde{D}_4^\mu$ correspond to the muon line and do not enter the electron-line corrections. As a last input to the LBK theorem~\eqref{eq:lbk_allorder} we have calculated the non-radiative two-loop amplitude using the known expression for the heavy-quark form factor~\cite{Mastrolia:2003yz,Bonciani:2003ai,Bernreuther:2004ih,Gluza:2009yy}. The corresponding result is expressed in terms of harmonic polylogarithms~\cite{Remiddi:1999ew} that can be evaluated with the \texttt{Mathematica} code \texttt{HPL}~\cite{Maitre:2005uu,Maitre:2007kp}.

\begin{figure}
    \centering
    \subfloat[$1/\epsilon^2$ pole]{
        \includegraphics[width=.7\textwidth]{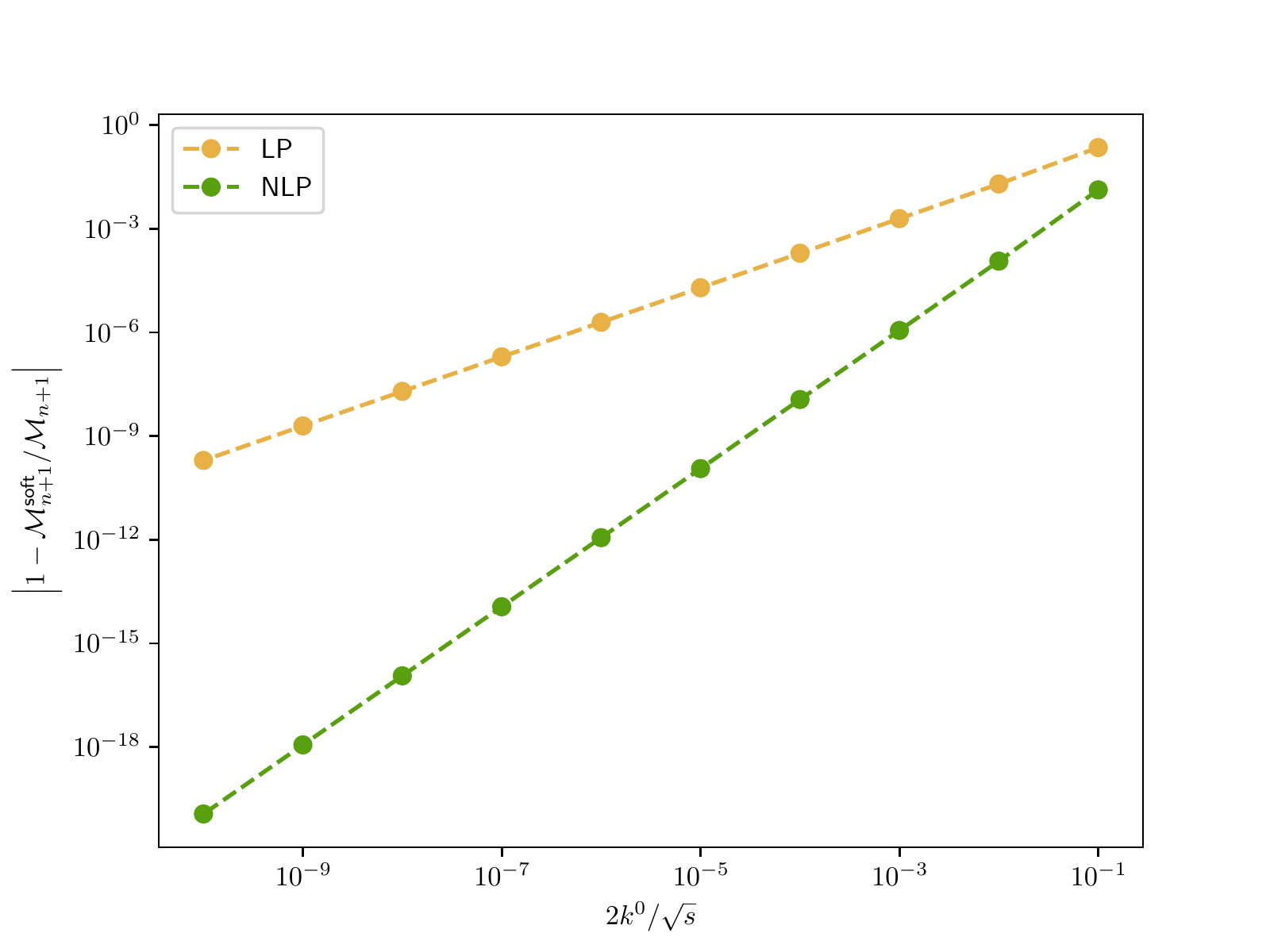}
    \label{fig:doublePole}
    } \\
    \subfloat[$1/\epsilon$ pole]{
        \includegraphics[width=.7\textwidth]{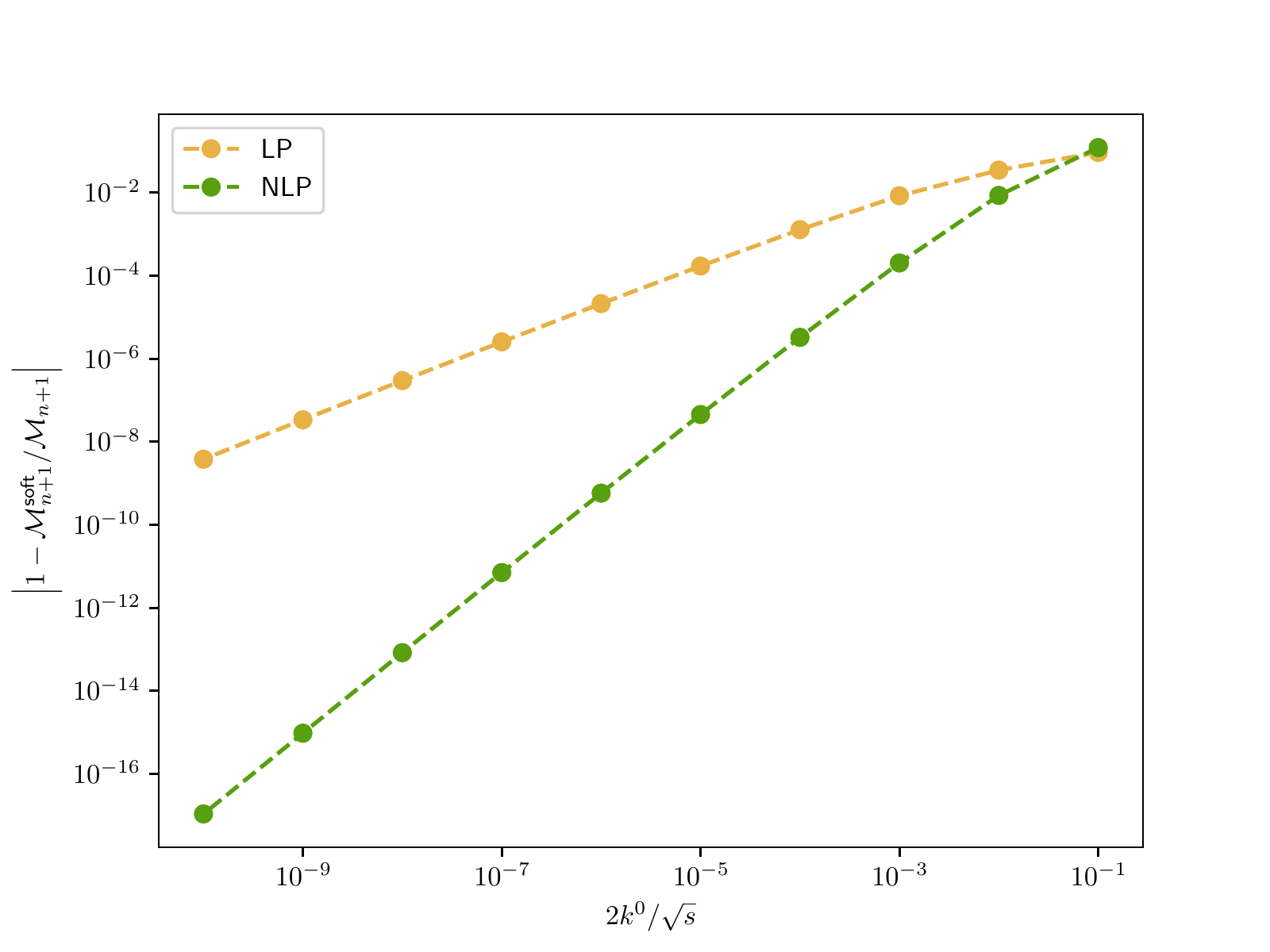}
    \label{fig:singlePole}
    }
\caption{The convergence of the IR poles of the soft approximation $\mathcal{M}_{n+1}^\text{soft}$ at LP and NLP as a function of the normalised photon energy $2 k^0/\sqrt{s}$. The exact reference values have been calculated using the YFS exponentiation formula. As expected, the relative difference of the soft approximation to the exact value tends to zero for decreasing photon energies $k^0 \to 0$. The inclusion of the NLP term gives a significant improvement of the approximation.}
\label{fig:muonePole}
\end{figure}

As a check of the obtained expression we compare the NLP expanded poles with the exact IR prediction. In fact, the exact poles can be calculated using the YFS exponentiation formula~\cite{Yennie:1961ad}
\begin{equation}
	e^{\hat{\mathcal{E}}} \mathcal{M}_{n+1} 
	= e^{\hat{\mathcal{E}}}  \sum_{\ell=0}^\infty \mathcal{M}_{n+1}^{(\ell)}
	= \text{finite}\, ,
\end{equation}
which yields for the singular two-loop part
\begin{equation}
	\mathcal{M}_{n+1}^{(2)}
	= - \hat{\mathcal{E}} \mathcal{M}_{n+1}^{(1)}
	- \frac{1}{2} \hat{\mathcal{E}} ^2  \mathcal{M}_{n+1}^{(0)}
	+ \mathcal{O}(\epsilon^0) \, .
\end{equation}
The explicit form of the integrated eikonal $\hat{\mathcal{E}}$ can be found in~\cite{Frederix:2009yq}. The diagrams entering the radiative tree-level and one-loop squared amplitudes, $\mathcal{M}_{n+1}^{(0)}$ and $\mathcal{M}_{n+1}^{(1)}$, have been generated with \texttt{QGraf}~\cite{NOGUEIRA1993279} and evaluated using \texttt{Package-X}~\cite{Patel:2015tea}.

Figure~\ref{fig:muonePole} shows the relative difference between the LP and NLP soft approximations and the exact poles as a function of the normalised photon energy $2 k^0 /\sqrt{s}$. The $1/\epsilon^2$ (Figure~\ref{fig:doublePole}) as well as the $1/\epsilon$ (Figure~\ref{fig:singlePole}) pole exhibit the expected behavior. In both cases the inclusion of the NLP term in the soft expansion results in a significant improvement of the approximation. This provides a non-trivial validation of the generalised LBK theorem~\eqref{eq:lbk_allorder}.

\section{Conclusion}
\label{sec:conclusion}

We have demonstrated that to NLP the soft limit of one-photon radiation is entirely determined by the corresponding non-radiative contribution for unpolarised scattering. This generalises the LBK theorem to all orders, which, in turn, can be viewed as an extension of the LP result of Yennie, Frautschi, and Suura. Working in the framework of HQET, we have proven the following two statements that form the basis for the all-order result.

First, beyond one loop, all soft virtual corrections vanish. As a result, the soft function calculated in~\cite{Engel:2021ccn} is one-loop exact. Second, the spin-flipping magnetic contributions cancel at the level of the unpolarised squared amplitude. The remaining hard matching corrections can be related to the non-radiative process following the original tree-level derivation of the LBK theorem. In fact, the same form of the LBK differential operator can be used. These results permit to generalise the one-loop theorem derived in~\cite{Engel:2021ccn} to all orders in perturbation theory.

As a first application, we have calculated the real-virtual-virtual electron-line corrections for muon-electron scattering to NLP in the soft limit. This new result is relevant in light of the MUonE experiment requiring high-precision theory predictions below 10 ppm, where QED corrections beyond NNLO become important. We have compared the singular part of the obtained expression with the exact IR poles predicted from eikonal exponentiation. This represents a non-trivial consistency check of our findings.

Other possible applications of the generalised LBK theorem are left for future studies. This includes the extension to multiple soft-photon radiation and its application to the resummation of large logarithms. Furthermore, the theorem can be used to cross check calculations of amplitudes making sure that the corresponding result correctly reproduces the NLP soft behaviour. Finally, in situations where a full computation is not feasible, it can serve as an approximation or a means to `bootstrap' the amplitude.

\subsection*{Acknowledgement} 

We are very grateful to Maximilian Stahlhofen as well as Wan-Li Ju for helpful discussions regarding heavy-quark effective theory and its application to the LBK theorem. Furthermore, we thank Robin Brüser, Mathieu Pellen, Marco Rocco, Adrian Signer, Maximilian Stahlhofen, and Yannick Ulrich for the careful reading of the manuscript as well as valuable suggestions for improvement. Finally, we would like to express our gratitude towards Adrian Signer and Yannick Ulrich whose collaboration has formed the basis for the present work. TE was supported by the German Federal Ministry for
Education and Research (BMBF) under contract no. 05H21VFCAA.

\newpage
\begin{appendix}
\label{sec:appendix}

\section{Eikonal identities}
\label{sec:eik_id}

The trivial Dirac structure of the eikonal vertex~\eqref{eq:eikonal_rules} allows us to derive so-called eikonal identities that yield a very compact structure for corrections with an arbitrary number of soft photons. In the following, we prove the two eikonal identities given in~\eqref{eq:eikid_conv} and~\eqref{eq:eikid_gen} that are used in Section~\ref{sec:soft} to evaluate the NLP amplitude to all orders.

\subsection{On-shell eikonal identity}
\label{sec:eikid_conv}

The conventional on-shell eikonal identity is given by
\begin{equation}
\label{eq:eikid_conv}
    	\mathcal{R}_{n} 
    	= \begin{tikzpicture}[scale=.8,baseline={(0,-.1)}]
       	    
   \draw[line width=.3mm,style=double]  (-1.2,0) -- (1,0);

   \node at (-1.1,.3) {\footnotesize{$v$}};

   \draw[line width=0.3mm,photon]  (-.2,0) -- (-1.2,.95) node[left]{\footnotesize{$p_1$}};
   \draw[line width=0.3mm,photon]  (-.2,0) -- (.8,.95) node[right]{\footnotesize{$p_n$}};
   \draw[line width=.3mm]  [fill=black] (-.2,.8) circle (0.02);
   \draw[line width=.3mm]  [fill=black] (-.4,.75) circle (0.02);
   \draw[line width=.3mm]  [fill=black] (0.,.75) circle (0.02);

   \draw[line width=.3mm]  [fill=white] (-.2,0) circle (0.2);
   \draw[line width=.3mm] (-.2-.15,0.-.15) -- (-.2+.15,0.+.15);
   \draw[line width=.3mm] (-.2+.15,0.-.15) -- (-.2-.15,0.+.15);

   \draw[line width=.3mm]  [fill=stuff] (1,0) circle (0.4);
       	\end{tikzpicture}
     	\equiv \sum_{\sigma}
       	 \begin{tikzpicture}[scale=.8,baseline={(0,-.1)}]
       	    
   \draw[line width=.3mm,style=double]  (-1.2,0) -- (1,0);

   \node at (-1.1,.3) {\footnotesize{$v$}};

   \draw[line width=0.3mm,photon]  (-.7,0) -- (-.7,1.) node[left]{\footnotesize{$p_{\sigma(1)}$}};
   \draw[line width=0.3mm,photon]  (.4,0) -- (.4,1.) node[right]{\footnotesize{$p_{\sigma(n)}$}};
   \draw[line width=.3mm]  [fill=black] (.1,.4) circle (0.02);
   \draw[line width=.3mm]  [fill=black] (-.15,.4) circle (0.02);
   \draw[line width=.3mm]  [fill=black] (-.4,.4) circle (0.02);

   \draw[line width=.3mm]  [fill=stuff] (1,0) circle (0.4);
       	\end{tikzpicture}
    	= \prod_{i=1}^n \frac{Q v^{\mu_i}}{p_i \cdot v} \, .
\end{equation}
The crossed vertex represents all possible attachments of the photons, which, in turn, can be written as the sum over all possible permutations $\sigma$ of the $n$ incoming momenta $p_i$. 

To prove this we consider the single term
\begin{subequations}
\label{eq:specificperm}
\begin{align}
	\mathcal{R}^{q,r}_n	
    	&= \begin{tikzpicture}[scale=.8,baseline={(0,-.1)}]
       	    
   \draw[line width=.3mm,style=double]  (-2.3,0) -- (1,0);

   \node at (-2.2,.3) {\footnotesize{$v$}};

   \draw[line width=0.3mm,photon]  (-1.8,0) -- (-1.8,1.);
   \draw[line width=.3mm]  [fill=black] (-1.6,.4) circle (0.02);
   \draw[line width=.3mm]  [fill=black] (-1.5,.4) circle (0.02);
   \draw[line width=.3mm]  [fill=black] (-1.4,.4) circle (0.02);
   \draw[line width=0.3mm,photon]  (-1.2,0) -- (-1.2,1.);
   \draw [decorate, decoration = {brace},line width = 0.3mm] (-1.9,1.1) --  (-1.1,1.1);
   \draw (-1.5,1.4) node[] {\footnotesize{$q$}};

   \draw[line width=0.3mm,photon]  (.4,0) -- (.4,1.);
   \draw[line width=.3mm]  [fill=black] (.2,.4) circle (0.02);
   \draw[line width=.3mm]  [fill=black] (.1,.4) circle (0.02);
   \draw[line width=.3mm]  [fill=black] (0.,.4) circle (0.02);
   \draw[line width=0.3mm,photon]  (-.2,0) -- (-.2,1.);
   \draw [decorate, decoration = {brace},line width = 0.3mm] (-.3,1.1) --  (.5,1.1);
   \draw (.1,1.4) node[] {\footnotesize{$r$}};

   \draw[line width=0.3mm,photon]  (-.7,0) -- (-.7,1.);
   \node at (-.7,1.2) {\footnotesize{$p_n$}};

   \draw[line width=.3mm]  [fill=stuff] (1,0) circle (0.4);
       	\end{tikzpicture}
	\\
	&= \Bigg( \prod_{i=1}^q \frac{Q v^{\mu_i}}{\sum_{j=1}^i p_j \cdot v}  \Bigg)
	\frac{Q v^{\mu_n}}{\sum_{j=1}^q p_j \cdot v + p_n \cdot v}
	\Bigg( \prod_{i=q+1}^{q+r} \frac{Q v^{\mu_i}}{\sum_{j=1}^i p_j \cdot v + p_n \cdot v} \Bigg)
\end{align}
\end{subequations}
in the permutation sum of~\eqref{eq:eikid_conv} with $q+r=n-1$. If $q > 0$ we apply the partial fraction decomposition
\begin{equation}
	\frac{1}{\sum_{j} p_j \cdot v} \,
	\frac{1}{\sum_{j} p_j \cdot v + p_n \cdot v}
	= -\frac{1}{p_n \cdot v} \Bigg(
	 \frac{1}{\sum_{j} p_j \cdot v + p_n \cdot v}
	 -\frac{1}{\sum_{j} p_j \cdot v}
	  \Bigg)
\end{equation}
to rewrite~\eqref{eq:specificperm} as
\begin{equation}
	\mathcal{R}^{q,r}_n
	= \tilde{\mathcal{R}}^{q,r}_n-\tilde{\mathcal{R}}^{q+1,r-1}_n
\end{equation}
with
\begin{equation}
	\tilde{\mathcal{R}}^{q,r\geq 0}_n
	= - \frac{Q v^{\mu_n}}{p_n \cdot v}
	\prod_{i=1}^{q-1} \frac{Q v^{\mu_i}}{\sum_{j=1}^i p_j \cdot v} 
	\prod_{i=q}^{q+r} \frac{Q v^{\mu_i}}{\sum_{j=1}^i p_j \cdot v + p_n \cdot v}
\end{equation}
and
\begin{equation}
	\tilde{\mathcal{R}}^{q+1,-1}_n
	= - \frac{Q v^{\mu_n}}{p_n \cdot v}
	\prod_{i=1}^{q} \frac{Q v^{\mu_i}}{\sum_{j=1}^i p_j \cdot v} \, .
\end{equation}
The special case $q=0$ of~\eqref{eq:specificperm} is given by
\begin{align}
	\mathcal{R}^{0,r}_n = -\tilde{R}^{1,r-1}_n \, .
\end{align}
Summing over all possible insertions of $p_n$ yields the telescoping series
\begin{subequations}
\label{eq:insertsum}
\begin{align}
	\sum_{q+r=n-1} \mathcal{R}^{q,r}_n
	&=  \mathcal{R}^{0,n-1}_n +  \sum_{q=1}^{n-1} \mathcal{R}^{q,n-1-q}_n
	\\
	&= - \tilde{\mathcal{R}}^{1,n-2}_n 
	+ \sum_{q=1}^{n-1} \Big( \tilde{\mathcal{R}}^{q,n-1-q}_n -  \tilde{\mathcal{R}}^{q+1,n-1-q-1}_n \Big)
	\\
	&= -\mathcal{\tilde{R}}^{n,-1}_n
	\\
	&= \frac{Q v^{\mu_n}}{p_n \cdot v}
	\begin{tikzpicture}[scale=.8,baseline={(0,-.1)}]
           
   \draw[line width=.3mm,style=double]  (-1.2,0) -- (1,0);

   \node at (-1.2,.3) {\footnotesize{$v$}};

   \draw[line width=0.3mm,photon]  (-.7,0) -- (-.7,1.) node[left]{\footnotesize{$p_1$}};
   \draw[line width=0.3mm,photon]  (.4,0) -- (.4,1.) node[right]{\footnotesize{$p_{n-1}$}};
   \draw[line width=.3mm]  [fill=black] (.1,.4) circle (0.02);
   \draw[line width=.3mm]  [fill=black] (-.15,.4) circle (0.02);
   \draw[line width=.3mm]  [fill=black] (-.4,.4) circle (0.02);

   \draw[line width=.3mm]  [fill=stuff] (1,0) circle (0.4);
       \end{tikzpicture} \, .
\end{align}
\end{subequations}
Hence, we arrive at the relation
\begin{equation}
	\mathcal{R}_n = \frac{Q v^{\mu_n}}{p_n \cdot v} \mathcal{R}_{n-1} \, ,
\end{equation}
which recursively applied proves~\eqref{eq:eikid_conv}.

\subsection{Off-shell eikonal identity}
\label{sec:eikid_gen}

In the case of the generalised eikonal identity we consider $n$ photons with momenta $\ell_j$ that connect to an off-shell heavy-fermion propagator with residual momentum $p \neq 0$. The off-shellness is indicated in the diagram~\eqref{eq:eikid_gen} below by the small gray blob on the left side of the heavy-fermion propagator. We further restrict ourselves to processes with a single external soft photon, which is the scenario considered in this work. The identity given below therefore only applies to this specific case. The only soft scale is thus the photon momentum $k$. It either enters via the off-shellness $p \cdot v$ or corresponds to one of the photons $\ell_j$. All other photons are virtual particles that are part of a loop. The identity is then given by
\begin{equation}
\label{eq:eikid_gen}
	\mathcal{T}_{n}^{(p)}
    	= \begin{tikzpicture}[scale=.8,baseline={(0,-.1)}]
       	    
   \draw[line width=.3mm,style=double]  (-1.5,0) -- (1,0);

   \draw[line width=0.3mm,photon]  (-.25,0) -- (-1.25,.95) node[left]{\footnotesize{$\ell_1$}};
   \draw[line width=0.3mm,photon]  (-.25,0) -- (.75,.95) node[right]{\footnotesize{$\ell_n$}};
   \draw[line width=.3mm]  [fill=black] (-.25,.8) circle (0.02);
   \draw[line width=.3mm]  [fill=black] (-.45,.75) circle (0.02);
   \draw[line width=.3mm]  [fill=black] (-.05,.75) circle (0.02);

   \draw[line width=.3mm]  [fill=white] (-.25,0) circle (0.2);
   \draw[line width=.3mm] (-.25-.15,0.-.15) -- (-.25+.15,0.+.15);
   \draw[line width=.3mm] (-.25+.15,0.-.15) -- (-.25-.15,0.+.15);

   \draw[line width=.3mm]  [fill=stuff] (1.25,0) circle (0.4);
   \draw[line width=.3mm]  [fill=stuff] (-1.5,0) circle (0.15);
   \node at (-1.5,.4) {\footnotesize{$v,p$}};
       	\end{tikzpicture}
	\equiv \sum_{\sigma}
       	 \begin{tikzpicture}[scale=.8,baseline={(0,-.1)}]
       	    \draw[line width=.3mm,style=double]  (-1.5,0) -- (1,0);

   \draw[line width=0.3mm,photon]  (-.7,0) -- (-.7,1.) node[left]{\footnotesize{$\ell_{\sigma(1)}$}};
   \draw[line width=0.3mm,photon]  (.4,0) -- (.4,1.) node[right]{\footnotesize{$\ell_{\sigma(n)}$}};
   \draw[line width=.3mm]  [fill=black] (.1,.4) circle (0.02);
   \draw[line width=.3mm]  [fill=black] (-.15,.4) circle (0.02);
   \draw[line width=.3mm]  [fill=black] (-.4,.4) circle (0.02);

   \draw[line width=.3mm]  [fill=stuff] (1.25,0) circle (0.4);
   \draw[line width=.3mm]  [fill=stuff] (-1.5,0) circle (0.15);
   \node at (-1.5,.4) {\footnotesize{$v,p$}};
       	\end{tikzpicture}
     	= \Bigg( \prod_{i=1}^n \frac{-s_i Q v^{\mu_i}}{s_i \ell_i \cdot v+i0} \Bigg)
       	 \frac{i}{\widetilde{\sum}_j \ell_j \cdot v + \tilde{p} \cdot v} \, ,
\end{equation}
which is valid up to scaleless loop integrals. Due to the off-shellness we now have one propagator more than in the on-shell formula~\eqref{eq:eikid_conv}. Each open photon line has one of the following four roles in the full diagram: It (i) attaches to the external leg labelled by $v$, (ii) corresponds to the external photon, (iii) attaches to a different external leg, or (iv) connects to another open photon line to form a loop. The momentum flow in~\eqref{eq:eikid_gen} is either incoming or outgoing depending on where the corresponding photon attaches. A momentum $\ell_i$ is outgoing if it connects to the external leg labelled by $v$. In the case it attaches to a different external line or corresponds to the external photon it is considered to be incoming. If two $\ell_i$ form a loop, one momentum is taken to be incoming and the other one outgoing. In this case the modified sum $\widetilde{\sum}$ only takes into account the incoming momentum. The reason for this will become clear in the proof below. Furthermore, there is an implicit $\ell_i$ dependence hidden in $p$ if the photon connects to the external leg labelled by $v$. Removing this implicit dependence yields $\tilde{p}$. Finally note that $s_i$ is $+1$ if $\ell_i$ is incoming and $-1$ otherwise. We explicitly display the causal $+i0$ prescription to show that these sign factors do not cancel.

To prove this identity we follow a similar approach as for the on-shell case discussed in Section~\ref{sec:eikid_conv}. We consider the contribution
\begin{subequations}
\label{eq:specificperm_gen}
\begin{align}
	\mathcal{T}^{q,r}_n	
    	&= \begin{tikzpicture}[scale=.8,baseline={(0,-.1)}]
       	    
   \draw[line width=.3mm,style=double]  (-2.3,0) -- (1,0);

   \draw[line width=0.3mm,photon]  (-1.8,0) -- (-1.8,1.);
   \draw[line width=.3mm]  [fill=black] (-1.6,.4) circle (0.02);
   \draw[line width=.3mm]  [fill=black] (-1.5,.4) circle (0.02);
   \draw[line width=.3mm]  [fill=black] (-1.4,.4) circle (0.02);
   \draw[line width=0.3mm,photon]  (-1.2,0) -- (-1.2,1.);
   \draw [decorate, decoration = {brace},line width = 0.3mm] (-1.9,1.1) --  (-1.1,1.1);
   \draw (-1.5,1.4) node[] {\footnotesize{$q$}};

   \draw[line width=0.3mm,photon]  (.4,0) -- (.4,1.);
   \draw[line width=.3mm]  [fill=black] (.2,.4) circle (0.02);
   \draw[line width=.3mm]  [fill=black] (.1,.4) circle (0.02);
   \draw[line width=.3mm]  [fill=black] (0.,.4) circle (0.02);
   \draw[line width=0.3mm,photon]  (-.2,0) -- (-.2,1.);
   \draw [decorate, decoration = {brace},line width = 0.3mm] (-.3,1.1) --  (.5,1.1);
   \draw (.1,1.4) node[] {\footnotesize{$r$}};

   \draw[line width=0.3mm,photon]  (-.7,0) -- (-.7,1.);
   \node at (-.7,1.2) {\footnotesize{$\ell_n$}};

   \draw[line width=.3mm]  [fill=stuff] (1,0) circle (0.4);
   \draw[line width=.3mm]  [fill=stuff] (-2.3,0) circle (0.15);
   \node at (-2.3,.4) {\footnotesize{$v,p$}};
       	\end{tikzpicture}
	\\
	&= \frac{i}{p \cdot v}
	\Bigg( \prod_{i=1}^q \frac{Q v^{\mu_i}}{\sum_{j=1}^i \ell_j \cdot v + p \cdot v}  \Bigg)
	\frac{Q v^{\mu_n}}{\sum_{j=1}^q \ell_j \cdot v + \ell_n \cdot v+ p \cdot v}
	\Bigg( \prod_{i=q+1}^{q+r} \frac{Q v^{\mu_i}}{\sum_{j=1}^i \ell_j \cdot v + \ell_n \cdot v+ p \cdot v} \Bigg)
\end{align}
\end{subequations}
in the permutation sum~\eqref{eq:eikid_gen} with $q+r=n-1$. If $q>0$ we apply the partial fraction decomposition
\begin{align}
	\frac{1}{\sum_{j} \ell_j \cdot v + p \cdot v}\, 
	\frac{1}{\sum_{j} \ell_j \cdot v + \ell_n \cdot v + p \cdot v}
	= -\frac{1}{\ell_n \cdot v} \Bigg(&
	 \frac{1}{\sum_{j} \ell_j \cdot v + \ell_n \cdot v + p \cdot v}
	-\frac{1}{\sum_{j} \ell_j \cdot v + p \cdot v} \Bigg)
\end{align}
to rewrite~\eqref{eq:specificperm_gen} as
\begin{equation}
	\mathcal{T}_{n}^{q,r}
    	= \tilde{\mathcal{T}}^{q,r}_n-\tilde{\mathcal{T}}^{q+1,r-1}_n
\end{equation}
with
\begin{equation}
	\tilde{\mathcal{T}}^{q,r\geq0}_n
	= - \frac{i}{p \cdot v}
	\frac{Q v^{\mu_n}}{\ell_n \cdot v}
	\prod_{i=1}^{q-1} \frac{Q v^{\mu_i}}{\sum_{j=1}^i \ell_j \cdot v + p \cdot v} 
	\prod_{i=q}^{q+r} \frac{Q v^{\mu_i}}{\sum_{j=1}^i \ell_j \cdot v + \ell_n \cdot v + p \cdot v}
\end{equation}
and
\begin{equation}
	\tilde{\mathcal{T}}^{q+1,-1}_n
	= - \frac{i}{p \cdot v}
	\frac{Q v^{\mu_n}}{\ell_n \cdot v}
	\prod_{i=1}^{q} \frac{Q v^{\mu_i}}{\sum_{j=1}^i \ell_j \cdot v + p \cdot v} \, .
\end{equation}
The special case $q=0$ of~\eqref{eq:specificperm_gen} is given by
\begin{equation}
	\mathcal{T}^{0,r}_n
	= -\frac{\ell_n \cdot v}{\ell_n \cdot v + p \cdot v} \tilde{\mathcal{T}}^{1,r-1}_n \, .
\end{equation}
Contrary to~\eqref{eq:insertsum}, this term does not cancel when summing over all insertions. Instead, we find
\begin{subequations}
\begin{align}
	\sum_{q+r=n-1} \mathcal{T}^{q,r}_n
	&=  \mathcal{T}^{0,n-1}_n +  \sum_{q=1}^{n-1} \mathcal{T}^{q,n-1-q}_n
	\\
	&= -\frac{\ell_n \cdot v}{\ell_n \cdot v + p \cdot v}\tilde{\mathcal{T}}^{1,n-2}_n 
	+ \sum_{q=1}^{n-1} \Big( \tilde{\mathcal{T}}^{q,n-1-q}_n -  \tilde{\mathcal{T}}^{q+1,n-1-q-1}_n \Big)
	\\
	&= \Big( -\frac{\ell_n \cdot v}{\ell_n \cdot v + p \cdot v} + 1 \Big) \tilde{\mathcal{T}}^{1,n-2}_n-\mathcal{\tilde{T}}^{n,-1}_n \, ,
\end{align}
\end{subequations}
which can be rewritten as
\begin{subequations}
\begin{align}
	\sum_{q+r=n-1} \mathcal{T}^{q,r}_n
	&=\frac{p \cdot v}{\ell_n \cdot v + p \cdot v} \tilde{\mathcal{T}}^{1,n-2}_n-\mathcal{\tilde{T}}^{n,-1}_n
	\\
	&= \frac{-Q v^{\mu_n}}{\ell_n \cdot v}
	\Bigg(
        \begin{tikzpicture}[scale=.8,baseline={(0,-.1)}]
           
   \draw[line width=.3mm,style=double]  (-1.4,0) -- (1,0);

   \draw[line width=0.3mm,photon]  (-.7,0) -- (-.7,1.) node[left]{\footnotesize{$\ell_1$}};
   \draw[line width=0.3mm,photon]  (.4,0) -- (.4,1.) node[right]{\footnotesize{$\ell_{n-1}$}};
   \draw[line width=.3mm]  [fill=black] (.1,.4) circle (0.02);
   \draw[line width=.3mm]  [fill=black] (-.15,.4) circle (0.02);
   \draw[line width=.3mm]  [fill=black] (-.4,.4) circle (0.02);

   \draw[line width=.3mm]  [fill=stuff] (1,0) circle (0.4);
   \draw[line width=.3mm]  [fill=stuff] (-1.4,0) circle (0.15);
   \node at (-1.55,.4) {\footnotesize{$v,p+\ell_n$}};
        \end{tikzpicture}
        -
	\begin{tikzpicture}[scale=.8,baseline={(0,-.1)}]
           
   \draw[line width=.3mm,style=double]  (-1.4,0) -- (1,0);

   \draw[line width=0.3mm,photon]  (-.7,0) -- (-.7,1.) node[left]{\footnotesize{$\ell_1$}};
   \draw[line width=0.3mm,photon]  (.4,0) -- (.4,1.) node[right]{\footnotesize{$\ell_{n-1}$}};
   \draw[line width=.3mm]  [fill=black] (.1,.4) circle (0.02);
   \draw[line width=.3mm]  [fill=black] (-.15,.4) circle (0.02);
   \draw[line width=.3mm]  [fill=black] (-.4,.4) circle (0.02);

   \draw[line width=.3mm]  [fill=stuff] (1,0) circle (0.4);
   \draw[line width=.3mm]  [fill=stuff] (-1.4,0) circle (0.15);
   \node at (-1.4,.4) {\footnotesize{$v,p$}};
        \end{tikzpicture}
        \Bigg) \, .
\end{align}
\end{subequations}
This implies the recursion relation
\begin{equation}
	\label{eq:recursion_general}
	\mathcal{T}_{n}^{(p)}
	= \frac{-Q v^{\mu_n}}{\ell_n \cdot v} \big( \mathcal{T}_{n-1}^{(p+\ell_n)}- \mathcal{T}_{n-1}^{(p)} \big) \, .
\end{equation}

At this point we have to consider the aforementioned four cases (i)-(iv) separately:
\begin{enumerate}[(i)]
	\item
	Since $\ell_n$ attaches to a different external leg, $p$ and therefore also $ \mathcal{T}_{n-1}^{(p)} $ is independent of $\ell_n$. Hence, the corresponding contribution in~\eqref{eq:recursion_general} is scaleless upon analytic integration over $\ell_n$ and we find the simplified recursion relation
	\begin{equation}
		\label{eq:eikgen_recursion1}
		\mathcal{T}_{n}^{(p)}
		=   \frac{-Q v^{\mu_n}}{\ell_n \cdot v}  \mathcal{T}_{n-1}^{(p+\ell_n)} \, .
	\end{equation}
	This relies on the aforementioned assumption that the open end of the $\ell_n$ line attaches to an external leg. As a consequence of the on-shell eikonal identity~\eqref{eq:eikid_conv} the corresponding loop propagators do not introduce a scale. 
	\item
	If $\ell_n = k$ and $n>1$ all loop integrals in $\mathcal{T}_{n-1}^{(p)}$ are scaleless. This follows from $\mathcal{T}_{n-1}^{(p)}$ being independent of $k$ which is the only scale in the process. Furthermore, also for $n=1$ the contribution due to $\mathcal{T}_{0}^{(p)}$ can be neglected. Since by assumption $p \neq 0$, there is at least one scaleless loop integral present also in this case. As a result, the recursion relation~\eqref{eq:eikgen_recursion1} applies here as well.
	\item
	Since $\ell_n$ connects to the external leg labelled by $v$, we have an implicit $\ell_n$ dependence in $p$. Making this explicit by redefining $p \to \tilde{p}-\ell_n$ we find
	\begin{equation}
		\mathcal{T}_{n}^{(p)}
		= \frac{-Q v^{\mu_n}}{\ell_n \cdot v} \big(  \mathcal{T}_{n-1}^{(\tilde{p})}- \mathcal{T}_{n-1}^{(\tilde{p}-\ell_n)}\big)
		=  \frac{Q v^{\mu_n}}{\ell_n \cdot v}  \mathcal{T}_{n-1}^{(\tilde{p}-\ell_n)} \, ,
	\end{equation}
	where we can neglect the scaleless contribution corresponding to $\mathcal{T}_{n-1}^{(\tilde{p})}$. Considering the loop momentum as outgoing, \emph{i.e.}\ replacing $\ell_n \to -\ell_n$, we find the recursion relation
	\begin{equation}
	\label{eq:eikgen_recursion2}
		\mathcal{T}_{n}^{(p)}
		=  \frac{Q v^{\mu_n}}{-\ell_n \cdot v + i0}  \mathcal{T}_{n-1}^{(\tilde{p}+\ell_n)} \, .
	\end{equation}
	The $+i0$ prescription is given explicitly here to emphasize that the minus sign in the propagator has a different meaning than the one in the numerator of~\eqref{eq:eikgen_recursion1}.
	\item
	In this case $\ell_n$ connects with another photon line $\ell_j$ to form a loop and therefore satisfies $\ell_j = - \ell_n$. Without loss of generality we can assume $j=n-1$. Applying~\eqref{eq:recursion_general} twice yields
	\begin{equation}
		\mathcal{T}_{n}^{(p)}
		= \frac{1}{2} \frac{-Q v^{\mu_n}}{\ell_n \cdot v} \frac{-Q v^{\mu_{n-1}}}{-\ell_n \cdot v}
		\big( \mathcal{T}_{n-2}^{(p)}
		- \mathcal{T}_{n-2}^{(p+\ell_n)}
		- \mathcal{T}_{n-2}^{(p-\ell_n)}
		+ \mathcal{T}_{n-2}^{(p)} \big) \, ,
	\end{equation}
	where the factor $1/2$ corrects for the double counting due to the indistinguishability of $\ell_n$ and $\ell_{n-1}$. Omitting the scaleless contributions and shifting $\ell_n \to -\ell_n$ in the term corresponding to $\mathcal{T}_{n-2}^{(p-\ell_n)}$, we find
	\begin{equation}
	\label{eq:eikgen_recursion3}
		\mathcal{T}_{n}^{(p)}
		= \frac{-Q v^{\mu_n}}{\ell_n \cdot v} \frac{Q v^{\mu_{n-1}}}{\ell_{n-1} \cdot v}  \mathcal{T}_{n-2}^{(p+\ell_n)} \, .
	\end{equation}
	This can be brought to a form consistent with the previous cases by taking $\ell_{n-1}$ as outgoing, which yields
	\begin{equation}
	\label{eq:eikgen_recursion3}
		\mathcal{T}_{n}^{(p)}
		= \frac{-Q v^{\mu_n}}{\ell_n \cdot v+i0} \frac{Q v^{\mu_{n-1}}}{-\ell_{n-1} \cdot v+i0}  \mathcal{T}_{n-2}^{(p+\ell_n)} \, .
	\end{equation}
	Again, we explicitly display the $+i0$ prescription of the propagators to indicate the difference for incoming and outgoing momenta.
\end{enumerate}

The recursive application of the three relations~\eqref{eq:eikgen_recursion1},  \eqref{eq:eikgen_recursion2}, and \eqref{eq:eikgen_recursion3} terminates at $\mathcal{T}_0^{\big(\widetilde{\sum}_j \ell_j+\tilde{p}\big)}$ and proves the off-shell eikonal identity~\eqref{eq:eikid_gen}.

\end{appendix}

\bibliographystyle{JHEP}
\bibliography{lbk.bib}

\end{document}